\documentclass[runningheads,10pt]{llncs}
\usepackage[letterpaper, margin=1in]{geometry}
\usepackage[T1]{fontenc}
\usepackage{graphicx}
\usepackage{amsmath,amssymb,mathtools,bbm,mathrsfs}

\usepackage{subcaption, rotating}
\usepackage{ulem,cancel}
\usepackage{xspace}
\usepackage{enumitem}
\usepackage{booktabs, multirow}
\usepackage{array}
\usepackage{float}
\usepackage{soul}

\usepackage[algo2e,ruled]{algorithm2e}
\usepackage{tikz}
\usetikzlibrary{calc}
\usetikzlibrary{positioning, quotes, arrows.meta, backgrounds, fit, shapes, shadings, 3d}
\usepackage{adjustbox}

\tikzset{
    3Dstack/.style={
        x={(1cm, 0cm)},
        y={(0cm, 1cm)},
        z={(2mm, 5mm)},
    },
}

\tikzset{
    ncbar angle/.initial=90,
    ncbar/.style={
        to path=(\tikztostart)
        -- ($(\tikztostart)!#1!\pgfkeysvalueof{/tikz/ncbar angle}:(\tikztotarget)$)
        -- ($(\tikztotarget)!($(\tikztostart)!#1!\pgfkeysvalueof{/tikz/ncbar angle}:(\tikztotarget)$)!\pgfkeysvalueof{/tikz/ncbar angle}:(\tikztostart)$)
        -- (\tikztotarget)
    },
    ncbar/.default=0.5cm,
}
 
\tikzset{square left brace/.style={ncbar=0.5cm}}
\tikzset{square right brace/.style={ncbar=-0.5cm}}
 
\tikzset{round left paren/.style={ncbar=0.5cm,out=120,in=-120}}
\tikzset{round right paren/.style={ncbar=0.5cm,out=60,in=-60}}

\newcommand{\defeq}{\coloneq}

\newcommand{\E}{\mathbb{E}}
\newcommand{\calP}{\mathcal{P}}
\DeclareMathOperator{\Pa}{\mathcal{P}}
\DeclareMathOperator{\scrP}{\mathscr{P}}

\newcommand{\Arm}{ARMADiLLO\@\xspace}

\newcommand{\A}{\mathscr{A}}
\newcommand{\bPa}{\boldsymbol{\Pa}}
\newcommand{\s}{\mathbf{s}}
\newcommand{\bu}{\mathbf{u}}
\newcommand{\bv}{\mathbf{v}}
\newcommand{\x}{\mathbf{x}}
\newcommand{\y}{\mathbf{y}}
\newcommand{\z}{\mathbf{z}}
\newcommand{\bP}{\mathbf{P}}
\newcommand{\bQ}{\mathbf{Q}}
\newcommand{\Q}{\mathbf{Q}}
\newcommand{\tQ}{\tilde{\mathbf{Q}}}

\newcommand{\xmap}{{\hat{\x}}_{\mathrm{MAP}}}
\newcommand{\xmm}{{\hat{\x}}_{\mathrm{MM}}} 

\begin{document}

\title{Algorithms for Reconstructing B Cell Lineages in the Presence of Context-Dependent Somatic Hypermutation}
\titlerunning{Lineage Reconstruction Under Context-Dependent Mutation}
%
\author{Yongkang Li\inst{1}\orcidID{0000-0003-1785-4689} 
\and 
Kevin Wiehe\inst{1}
\and Scott C. Schmidler\inst{1}\orcidID{0009-0006-3733-3716}}
\authorrunning{Y. Li et al.}
%
\institute{Duke University, Durham NC 27708, USA}
\maketitle              
\begin{abstract}
We introduce a method for approximating posterior probabilities of phylogenetic trees and reconstructing ancestral sequences under models of sequence evolution with site-dependence, where standard phylogenetic likelihood computations (pruning) fail. Our approach uses a combined data-augmentation and importance sampling scheme. A key advantage of our approach is the ability to leverage existing highly optimized phylogenetic software. 
We apply our approach to the reconstruction of B cell receptor affinity maturation lineages from high-throughput repertoire sequencing data and evaluate
the impact of incorporating site-dependence on the reconstruction accuracy of both trees and ancestral sequences. 
We show that accounting for context-dependence during inference always improves the estimates of both ancestral sequences and lineage trees on simulated datasets.
We also examine the impact 
of incorporating priors based on 
VDJ recombination models, and find that they
significantly improve ancestral sequence reconstruction in germline-encoded regions, but 
increase 
errors in non-templated nucleotides. We propose a modified, piecewise prior to address this 
and demonstrate that it improves empirical reconstruction accuracy. We apply our approach to the analysis of 
the 
HIV broadly neutralizing antibodies  
DH270 and CH235 which are important targets of current vaccine design efforts.\\
\textbf{Code availability:} 
\url{https://github.com/YongkangLi/PhyConD}


\end{abstract}

%
%
%


\newpage

\section{Introduction}

To prevent infectious diseases, the immune system must recognize and clear foreign pathogens capable of 
host infection. The adaptive 
immune system 
does so
using B cells 
with receptors specifically tailored to bind pathogens. Facing 
a near-infinite variety of 
pathogen protein surfaces, B cells 
rely on two diversification methods: 1) 
initial gene segment shuffling (VDJ recombination) to enable weak binding to the diversity of antigens encountered, and 2) a refinement step (affinity maturation) that substantially improves binding and 
BCR-antigen recognition specificity, accomplished via 
somatic hypermutation (SHM).  Understanding of 
this process 
and design of new vaccines
is greatly aided by computational and statistical 
analysis of data from high-throughput B cell repertoire sequencing. 
A key challenge is 
accurate inference of BCR lineages and unmutated common ancestors, and 
%
significant progress has been made on
these tools 
\cite{kepler_reconstructing_2013,Ralph:2016,Marcou2018,dhar_bayesian_2020}.  However, 
existing methods rely on conventional ``independent-site" mutation models from molecular phylogenetics 
and unlike mutational processes elsewhere in the genome, these models do not adequately capture the SHM process, which is mediated by the enzyme activation induced cytidine deaminase (AID).  
Critically, AID activity is \textit{sequence-context dependent}, targeting sites 
in certain favored motifs.
Unfortunately, incorporation of 
context-dependent mutation models 
makes standard phylogenetic calculations intractable.
Here we develop 
algorithms for reconstructing B cell lineages under 
context-dependent SHM models. 
We build on recent work on  approximating pairwise sequence evolution probabilities 
\cite{Mathews:2023b,Mathews:2025a,Mathews:2025b}, extending these 
in non-trivial ways to trees   
using 
Monte Carlo 
methods of data augmentation, importance sampling, and sequential Monte Carlo.  Notably, our approach can be implemented 
directly using 
 existing phylogenetics software. 

\section{Background}

\subsection{BnAbs, Sequential Vaccine Design, and B Cell Lineages}

The goal of vaccination is to induce antibodies – 
secreted versions of B cell receptors (BCRs) – that 
neutralize pathogens and 
protect against infection. In 
infection, 
successive rounds of mutation 
and selection for antigen affinity 
(\textit{affinity maturation})  enables  
B cells to recognize pathogens with specificity. However, rapidly mutating viruses (HIV, influenza, SARS-CoV-2) escape this B cell response 
by viral diversification. 
It is thus 
desired to elicit \textit{broadly neutralizing} antibodies (bnAbs),
but these are rarely elicited 
due to a variety of factors  \cite{Walker_Burton_2018}:
low precursor frequencies in the human B cell receptor repertoire \cite{Haynes_35962033,Jardine_27013733}, 
highly mutated,
and containing essential mutations 
that are highly
improbable,
creating acquisition bottlenecks \cite{Wiehe:2018,Saunders_Wiehe_2019,Mathews:2023b}. 
Traditional vaccine design strategies have proven ineffective at eliciting bnAbs, motivating 
new strategies 
using known bnAbs as templates to design immunogens 
to 
guide B cell maturation 
the 
bnAb~\cite{Haynes_2012,Jardine_2015,Haynes_35962033,Steichen_2019,Kwong_29768174}. 
\textit{Sequential prime-boosting} 
aims to first induce low-frequency bnAb 
precursors with a \textit{priming} immunogen, 
then mature 
them with 
additional 
\textit{boosting} immunogens,
and relies heavily on inferring the precursor BCR sequence 
(the unmutated common ancestor (UCA)) from observed, clonally-related sequences \cite{kepler_reconstructing_2013,Haynes_2012}.  More generally 
reconstruction of B cell \textit{lineage} - the set of affinity maturation pathways for clone members 
- 
provides 
insight into maturation bottlenecks and
also 
precursors to 
serve as intermediate vaccine targets.
%
Existing 
lineage reconstruction methods such as  Cloanalyst \cite{kepler_reconstructing_2013}, IgOR \cite{Marcou2018}, 
Partis \cite{Ralph:2016} and LinearHam \cite{dhar_bayesian_2020} combine phylogenetic inference with 
models of VDJ recombination,
but follow the vast majority of molecular phylogenetics in assuming that 
sequence positions evolve independently;
%
%
this assumption is critical to 
tractability of 
phylogenetic inference. While 
appealing in their simplicity and computational convenience, 
these independent-site models 
fail to capture the critical context-dependence of SHM in 
affinity maturation.

\subsection{Affinity Maturation of B Cell Receptors: Context-Dependent Somatic Hypermutation}
\label{sec:ContextDependentSHM}

Unlike genomic 
mutation, 
SHM is mediated by a single enzyme
AID, which
induces mismatch lesions in DNA - and  subsequent error-prone repair - that leads to point mutations in BCRs \cite{Teng2007}. 
Critically AID activity is 
\textit{sequence dependent}, targeting 
certain favored motifs (“hot spots”) and disfavoring others (“cold spots”). 

\textit{S5F model and \Arm:} 
\label{sec:ARM}
We 
previously developed 
\Arm \cite{Wiehe:2018} for simulating 
SHM using context-dependent mutation probabilities 
from the S5F 
model~\cite{Yaari2013},
which
captures 
context-dependence 
of 
SHM 
using context-dependent site-mutation and substitution 
probabilities derived from 
NGS data \cite{Yaari2013} for the central nucleotide of 
1,024 5-mer motifs, 
given its (2 upstream and 2  downstream)
neighboring bases.
The model parameterizes mutation probabilities 
as a product of mutability scores and  substitution probabilities. 
\Arm 
has been used to 
forward-simulate SMH and
as a null model 
to test for selection \cite{Wiehe:2018,Tang:2023}. Unfortunately, this more realistic SHM model cannot be used in existing lineage reconstruction algorithms \cite{kepler_reconstructing_2013,Marcou2018,Ralph:2016,dhar_bayesian_2020}, 
as context-dependent mutation models directly 
violate the independent-site assumption
that makes standard phylogenetic calculations 
(pruning)
tractable. Since we might expect improved SHM models to 
provide improved reconstruction of UCAs and lineages, it is of great interest to develop methods for B cell lineage reconstruction which can handle context-dependent mutation models such as S5F/\Arm.


\section{Background}

\subsection{Standard Models of DNA Evolution}

Let $\x = (x_1, \ldots, x_\ell)$ and $\y = (y_1, \ldots, y_\ell)$ denote two nucleotide sequences of length $\ell$, where $x_i, y_i \in \A$ for $\A = \{\text{A,C,G,T} \}$. Standard phylogenetic models assume that each \( x_{i} \) evolves independently according to a time-homogeneous continuous-time Markov chain (CTMC) defined by rate matrices $\bQ^{(i)}$ with rates $\bQ^{(i)}_{ab} = \gamma^{(i)} ( b; a )$ where $\gamma^{(i)} : \A^2 \rightarrow ( 0, + \infty )$ specifies the rate at which site $i$ mutates from $a$ to $b$. 
The \textit{independent-site} assumption allows the straightforward calculation of sequence transition probabilities by 
\begin{equation}
\label{eqn:ISM}
p_{(T,\Q)}(\y \mid  \x) : =    \Pr(\x (t) = \y \mid \x (0) = \x) 
= \prod^{\ell}\nolimits_{i=1} p(y_{i} \mid x_{i},t) =  \prod^{\ell}\nolimits_{i=1} {(e^{tQ^{(i)}})}_{x_{i}y_{i}}
\end{equation}
where $\x (t) = (x_1(t),\ldots,x_n(t))$ denotes the state of the sequence at time $t$.

\subsection{Context-Dependent Mutation Models}

The independent site assumption, while computationally convenient, fails to capture a variety of known effects which induce dependence between sites; examples include include CpG di-nucleotide mutability \cite{Pederson:1998}, structural constraints in RNA and proteins \cite{Robinson:2003}, and enzyme-driven somatic hypermutation in B-cell affinity maturation \cite{Wiehe:2018,Mathews:2023b}.  A variety of models have been introduced to account for these various effects, most of which  can be written as \textit{context-dependent} rate matrices \cite{Mathews:2024}.
For nucleotide sequence $\x = (x_1,x_2,\ldots,x_\ell)$, let  $\tilde{x}_i = (x_{i-k}, \ldots, x_i, \ldots x_{i+k})$ denote the \textit{context} of site $x_i$ for $i \in \{1,\ldots,\ell\}$.
Relaxing the independent-site assumption, we can define a \textit{dependent site model} (DSM) with sites evolving under a time-\textit{inhomogeneous} CTMC and the rate at which $x_i$ mutates to $b \in \mathscr{A}$ in context $\Tilde{x}_i$ given by
%
\begin{equation}
    \Tilde{\gamma}^{(i)} ( b; {\Tilde{x}}_i ) := \phi ( b; {\Tilde{x}}_{i} ) \gamma^{(i)} ( b; x_i )
\label{Eqn:CDrates}
\end{equation}
%
for $\gamma_i$ 
the context-independent rate and $\phi: {\mathscr{A}}^{2k+1} \rightarrow ( 0, + \infty )$ 
a context-dependent multiplier;
then 
site $i$ exits $x_i$ 
at rate 
%
$\Tilde{\gamma}^{(i)} ( \cdot; {\Tilde{x}}_i ) := \sum_{b \in \mathscr{A} \setminus \{ x_i \}} \Tilde{\gamma}^{(i)} ( b; {\Tilde{x}}_i )$
%
and 
sequence $\x$ mutates 
at total rate ${\Tilde{\gamma}} ( \cdot; \x ) := \sum_{i=1}^{\ell} {\Tilde{\gamma}}^{(i)} ( \cdot; {\Tilde{x}}_{i} )$. Under DSMs, the calculation of sequence transition probabilities takes the form
\begin{equation}
\label{Eqn:DSM}
p_{(T,\tQ)}(\y \mid  \x) = {(e^{t \Tilde{\bQ}})}_{\x \y}
\end{equation}
for $\Tilde{\bQ}$ the $a^\ell \times a^\ell$ rate matrix 
of the CTMC 
on the space of all sequences defined by 
%
$\Tilde{\bQ}_{\x \x'} = 
\Tilde{\gamma}^{(i)} ( b; {\Tilde{x}}_i )$ if  $d_H ( \x, \x' ) = 1$ and $x_i \neq x_i' = b$,  
and 0 if $d_H(\x,\x')>1$.
%
For notational simplicity, all models are assumed to be position-independent, 
eliminating the site-superscript on rates and rate matrices.
Figure~\ref{fig:graphical_model}a
illustrates how, even when contexts are local, dependence propagates through time to link all sites along the sequence.
%
%
%
%
\Arm 
uses a 2-nearest-neighbor 5mer context, 
with 
each site $i$ 
assigned 
mutability score $m(\cdot; \tilde{x}_i)$ and 
substitution probability matrix $\eta(x^{\prime}_i;, \tilde{x}_i)$, 
to form 
context-dependent rates 
\eqref{Eqn:CDrates}. Mutability scores and 
substitution probabilities 
come from 
S5F 
\cite{Yaari2013} based on synonomous mutations in published NGS data. 

\subsection{Models of BCR Diversification for Lineage Reconstruction}

As noted 
VDJ recombination plays a critical role in diversification, generating the initial gene segment rearrangement upon which SMH
operates \cite{hozumi_evidence_1976,tonegawa_somatic_1983}. Probabilistic hidden Markov models (HMMs) have been successfully applied to model  VDJ recombination 
for 
clonal UCA reconstruction \cite{kepler_reconstructing_2013,dhar_bayesian_2020}. These 
methods marry an independent-site mutation model \eqref{eqn:ISM} for SHM with an HMM for stochastic VDJ shuffling. In Section~\ref{Sec:VDJPrior} below, we consider the incorporation of existing VDJ-models as prior distributions on the UCA,  with a likelihood calculated from our dependent-site SHM model, using the methods described in the next section.

\section{Methods} 
\label{sec:methods}

We introduce two algorithms for phylogenetic inference under context-dependent models of 
SMH.
We first consider inference based only on the observed clonal sequences (no VDJ recombination model); we refer to this as the \textit{pure phylogenetic model}. Incorporation of VDJ models is considered in Section~\ref{Sec:VDJPrior}. 


\subsection{Bayesian Lineage Reconstruction: the Pure Phylogenetic Model} 
\label{subsec:pure_phylo_model}

The posterior distribution  under the pure phylogenetic model is 
%
$\pi_{\text{PP}}(g,\Psi \mid Y) \propto p(Y \mid g,\Psi) f_G( g ) f_P(\Psi)$,
%
where $\Psi$ denotes 
parameters of the evolutionary model (rate matrices), 
$g$ 
the genealogy (
tree topology \& 
branch lengths),
and $Y = \{Y^1,\ldots,Y^k\}$ 
the set of observed sequences
$Y^i = (Y^i_1,\ldots Y^i_\ell)$. (For notational simplicity we assume 
sequences are 
equal length $\ell$, 
i.e. $Y$ is 
a multiple 
alignment). Here $f_P$ and $f_G$ are prior distributions
and \( p \left( Y \mid g, \Psi \right) \) 
the marginal likelihood of the observed sequences for a given tree $g$, the form of which
differs between the
ISM and 
DSM, with 
parameters denoted 
$\Psi_0$ and $\Psi_1$  respectively.

\subsection{Marginal likelihood calculation}
\label{Sec:MargLikCalc}

Evaluating 
$p(Y\mid g,\Psi)$ 
requires 
marginalizing 
over unobserved evolutionary histories of 
$Y$.
For $\bPa^{(T)} : [0, T] \rightarrow \mathscr{A}^{\ell} $ 
an evolutionary path over time 
$[0, T]$ 
and 
$\mathscr{P}^{(T)}$ 
the set of such paths, 
\eqref{Eqn:DSM} can be written \cite{nielsen_mapping_2002,rodrigue_uniformization_2008,Mathews:2024}:
\begin{equation}
p_{(T,\tQ)}(\y \mid  \x) = \int_{{\scrP}^{(T)}} p_{(T,\tQ)}( \y, {\bPa}^{(T)} \mid \x ) \nu(d\bPa).
\label{Eqn:DSMTransProb}
\end{equation}
%
Extending this 
to 
trees 
let 
$\bPa^{(g)} = {(\bPa^{(t_{e})})_{e \in E(g)}}$ be a \textit{tree-path}, 
for edge set $E(g)$ 
of $g$ and 
branch length $t_e$, 
and 
edge paths 
constrained to 
share internal node sequences. 
and let 
$\mathscr{P}^{(g)}$ be
the set of 
tree-paths. Then 
\begin{equation}
p( Y \mid g, \Psi) = 
\sum\nolimits_{\x} p ( Y, \x \mid g, \Psi ) = 
\sum\nolimits_{\x} \int_{{\mathscr{P}}^{(g)}} p ( Y, \bPa^{(g)}, \x \mid g, \Psi) d \bPa
\label{eq:MargLikPath}
\end{equation}
%
As noted 
calculating
\eqref{eq:MargLikPath}
by pruning is intractable 
for DSMs. 
Instead we can approximate 
by Monte Carlo.
The path representation 
has been used to sample 
$p(\Psi, \Pa^{(g)} \mid Y, g)$ 
using Markov chain Monte Carlo (MCMC)~\cite{nielsen_mapping_2002,Robinson:2003,bollback_simmap_2006,hobolth_sampling_2014}, 
but 
approximating 
marginal likelihoods from MCMC samples  is non-trivial~\cite{lartillot_computing_2006,arima_improved_2012,wolpert_stable_2012}, and determining 
MCMC runtimes 
difficult. 
A recent \cite{Mathews:2025a} 
importance sampling (IS) algorithm  
provides error bounds for approximating 
$p(\y\mid \x,t,\Psi_1)$ under DSMs 
via \eqref{Eqn:DSMTransProb}
with reweighted paths drawn from 
the ISM,
and \cite{Mathews:2025b} improves 
these bounds using sequential Monte Carlo (SMC).
Both
methods 
require only samples from the ISM, 
enabling the use of existing phylogenetics software. 
Here we consider two methods 
that extend 
the approaches in \cite{Mathews:2025a,Mathews:2025b} to trees to enable  
Monte Carlo approximation of 
\eqref{eq:MargLikPath} under arbritrary DSMs.
Both 
use \textit{data augmentation importance sampling}, 
sampling the augmented space of paths to 
enable evaluation of importance weights. The relative performance of these 
approaches is compared in Section~\ref{Sec:Results}.

\subsection{A stratified data augmentation importance sampler}
\label{sec:IS}

Our first approach leverages 
bounded-error approximation 
of 
edge transition probabilities \eqref{Eqn:DSMTransProb}  
\cite{Mathews:2025a}
by rewriting \eqref{eq:MargLikPath} as a marginalization over 
sequences at 
internal nodes of 
$g$.  
Let $\x =(x_1,\ldots,x_\ell)$ be the root 
UCA sequence 
and 
$Z = \{\z^1,\ldots,\z^r\}$ 
internal node sequences $\z^i = (z^i_1,\ldots,z^i_\ell)$, 
then
\begin{align} 
\label{eq:MargLik}
p( Y \mid g, \Psi) = 
\sum\nolimits_{\x, Z} p(Y, Z, \x \mid g, \Psi) & = 
\sum\nolimits_{\x, Z} \int_{\mathscr{P}^{(g)}} p ( Y, Z, \x, \bPa^{(g)} \mid g, \Psi) d \bPa, \\
& = \sum\nolimits_{\x, Z} p(Y, Z, \x \mid g, \Psi) \prod\nolimits_{(\bu,\bv, t) \in 
E(g)}
\int_{\mathscr{P}}
p( \bv, \bPa^{(t)} \mid \bu, \Psi) d \bPa 
\label{eq:MargLikEdgeProduct}
\end{align}
with
$E(g) := E(\x,Z,Y,g)$ 
the set of 
transitions along 
edges of $g$ for 
observed sequences $(\x,Z,Y)$.
The second line \eqref{eq:MargLikEdgeProduct} 
follows from the Markov property of the evolutionary model.

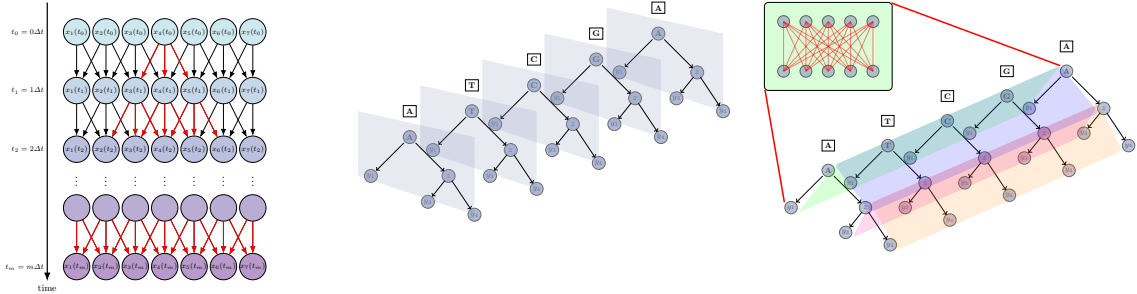
\begin{figure}[htbp]
\centering
\begin{subfigure}[t]{.32\textwidth}
\centering     \adjustbox{width=.68\linewidth, valign=T}{
\begin{tikzpicture}[
    >=Latex,
    x_node/.style={circle, draw, minimum size=0.9cm, inner sep=0pt, font=\scriptsize},
    time_label/.style={font=\scriptsize},
    time_arrow/.style={->, thick},
    dot_node/.style={font=\large},
    highlight/.style={->, thick, red}, 
]

\definecolor{c0}{RGB}{173,216,230}  
\definecolor{c1}{RGB}{158,188,218}
\definecolor{c2}{RGB}{140,150,198}
\definecolor{c3}{RGB}{138,118,182}
\definecolor{cm}{RGB}{136,86,167}   

\foreach \t/\y/\color in {0/0/c0, 1/-2/c1, 2/-4/c2, m/-8/cm} {

    \node[time_label, anchor=east] at (-4, \y) {$t_{\t}=\t\Delta t$};

    \foreach \i/\x in {1/-3, 2/-2, 3/-1, 4/0, 5/1, 6/2, 7/3} {
        \node[x_node, fill=\color!60] (x\i t\t) at (\x, \y) {$x_{\i}(t_{\t})$};
    }
}

\foreach \t/\y/\color in {3/-6/c3} {
    \foreach \i/\x in {1/-3, 2/-2, 3/-1, 4/0, 5/1, 6/2, 7/3} {
        \node[x_node, fill=\color!60] (x\i t\t) at (\x, \y) {};
    }
}

\foreach \x in {-3, -2, -1, 0, 1, 2, 3} {
    \node[dot_node] at (\x, -5) {$\vdots$};
}

\foreach \i/\prev/\t in {1/0/1, 1/1/2, 1/3/m, 7/0/1, 7/1/2, 7/3/m} {
    \draw[->, thick] (x\i t\prev) -- (x\i t\t);
}

\foreach \t/\prev in {1/0, 2/1, m/3} {
    \foreach \i in {2,3,4,5,6} {
        \pgfmathtruncatemacro{\im}{\i-1}
        \pgfmathtruncatemacro{\ip}{\i+1}
        \draw[->, thick] (x\im t\prev) -- (x\i t\t);
        \draw[->, thick] (x\i t\prev) -- (x\i t\t);
        \draw[->, thick] (x\ip t\prev) -- (x\i t\t);
    }
}

\foreach \prev/\t in {0/1, 1/2, 3/m} {
    \foreach \i in {2} {
        \pgfmathtruncatemacro{\im}{\i-1}
        \draw[->, thick] (x\i t\prev) -- (x\im t\t);
    }
}

\foreach \prev/\t in {0/1, 1/2, 3/m} {
    \foreach \i in {6} {
        \pgfmathtruncatemacro{\ip}{\i+1}
        \draw[->, thick] (x\i t\prev) -- (x\ip t\t);
    }
}

\foreach \src/\dst in {
    x4t0/x3t1, x4t0/x4t1, x4t0/x5t1,
    x3t1/x2t2, x3t1/x3t2, x3t1/x4t2,
    x4t1/x3t2, x4t1/x4t2, x4t1/x5t2,
    x5t1/x4t2, x5t1/x5t2, x5t1/x6t2,
    x2t3/x1tm, x2t3/x2tm, x2t3/x3tm,
    x3t3/x2tm, x3t3/x3tm, x3t3/x4tm,
    x4t3/x3tm, x4t3/x4tm, x4t3/x5tm,
    x5t3/x4tm, x5t3/x5tm, x5t3/x6tm,
    x6t3/x5tm, x6t3/x6tm, x6t3/x7tm,
    x1t3/x1tm, x1t3/x2tm, x7t3/x6tm, x7t3/x7tm} {
    \draw[highlight] (\src) -- (\dst);
}

\draw[very thick, ->] (-4, 1) -- (-4, -8.5) node[below] {time};

\end{tikzpicture}
}
\label{fig:nearest-neighbor-model}
\end{subfigure}
\begin{subfigure}[t]{.32\textwidth}
\centering
\adjustbox{width=.95\textwidth, valign=T}{
    \begin{tikzpicture}[3Dstack]
    \foreach \z/\base in {8/A, 6/G, 4/C, 2/T, 0/A} {
        \node[draw=black, fill=white, minimum width=0.5cm, minimum height=0.3cm] 
        at (\z, 1, \z) {\textbf{\base}}; 
        \ifnum\z=8 \definecolor{nodecolor}{rgb}{0.55,0.6,0.75} 
        \else\ifnum\z=6 \definecolor{nodecolor}{rgb}{0.55,0.6,0.75} 
        \else\ifnum\z=4 \definecolor{nodecolor}{rgb}{0.55,0.6,0.75} 
        \else\ifnum\z=2 \definecolor{nodecolor}{rgb}{0.55,0.6,0.75} 
        \else \definecolor{nodecolor}{rgb}{0.55,0.6,0.75}
        \fi\fi\fi\fi
     
        \node (root) at (\z,0,\z) {\textcolor{black}{\textbf{\base}}};
        \node (child1) at (\z-1.5,-1.5,\z) {\textcolor{black}{$y_{1}$}};
        \node (child2) at (\z+1.5,-1.5,\z) {\textcolor{black}{$z$}};
        \node (leaf3) at (\z+.7,-2.5,\z) {\textcolor{black}{$y_{3}$}};
        \node (leaf4) at (\z+2.5,-3,\z) {\textcolor{black}{$y_{4}$}};
        \draw[draw=black, fill=nodecolor, opacity=0.6] (root) circle (0.25cm);
        \draw[draw=black, fill=nodecolor, opacity=0.6] (child1) circle (0.25cm);
        \draw[draw=black, fill=nodecolor, opacity=0.6] (child2) circle (0.25cm);
        \draw[draw=black, fill=nodecolor, opacity=0.6] (leaf3) circle (0.25cm);
        \draw[draw=black, fill=nodecolor, opacity=0.6] (leaf4) circle (0.25cm);
      \begin{scope}
            \fill[nodecolor, opacity=0.2] (\z-2, -1.5, \z) -- (\z-2, 1, \z) -- (\z+2.5, 0, \z) -- (\z+2.5, -3, \z) -- cycle;
        \end{scope}
    \begin{scope}
            \fill[nodecolor, opacity=0.2] (\z, 0, \z) circle (0.25cm); 
            \fill[nodecolor, opacity=0.2] (\z-1.5, -1.5, \z) circle (0.25cm); 
            \fill[nodecolor, opacity=0.2] (\z+1.5, -1.5, \z) circle (0.25cm); 
            \fill[nodecolor, opacity=0.2] (\z+.7, -2.5, \z) circle (0.25cm); 
            \fill[nodecolor, opacity=0.2] (\z+2.5, -3, \z) circle (0.25cm); 
        \end{scope}
        \draw[->] (root) -- (child1);
        \draw[->] (root) -- (child2);
        \draw[->] (child2) -- (leaf3);
        \draw[->] (child2) -- (leaf4);
    }
    \end{tikzpicture}
}
\end{subfigure}
\begin{subfigure}{.32\textwidth}
\centering
\adjustbox{width=.95\textwidth, valign=T}{
    \begin{tikzpicture}[3Dstack, scale=0.85, every node/.style={scale=0.85}]

        \foreach \z/\base in {8/A, 6/G, 4/C, 2/T, 0/A} {
            \node[draw=black, fill=white, minimum width=0.5cm, minimum height=0.3cm] 
            at (\z, 1, \z) {\textbf{\base}}; 
            \ifnum\z=8 \definecolor{nodecolor}{rgb}{0.55,0.6,0.75} 
            \else\ifnum\z=6 \definecolor{nodecolor}{rgb}{0.7,0.6,0.75} 
            \else\ifnum\z=4 \definecolor{nodecolor}{rgb}{0.55,0.6,0.75} 
            \else\ifnum\z=2 \definecolor{nodecolor}{rgb}{0.7,0.6,0.75} 
            \else \definecolor{nodecolor}{rgb}{0.55,0.6,0.75}
            \fi\fi\fi\fi
        
            \node (root\z) at (\z,0,\z) {\textcolor{black}{\textbf{\base}}};
            \node (child1\z) at (\z-1.5,-1.5,\z) {\textcolor{black}{$y_{1}$}};
            \node (child2\z) at (\z+1.5,-1.5,\z) {\textcolor{black}{$z$}};
            \node (leaf3\z) at (\z+.7,-2.5,\z) {\textcolor{black}{$y_{3}$}};
            \node (leaf4\z) at (\z+2.5,-3,\z) {\textcolor{black}{$y_{4}$}};
            
            \draw[draw=black, fill=nodecolor, opacity=0.6] (root\z) circle (0.25cm);
            \draw[draw=black, fill=nodecolor, opacity=0.6] (child1\z) circle (0.25cm);
            \draw[draw=black, fill=nodecolor, opacity=0.6] (child2\z) circle (0.25cm);
            \draw[draw=black, fill=nodecolor, opacity=0.6] (leaf3\z) circle (0.25cm);
            \draw[draw=black, fill=nodecolor, opacity=0.6] (leaf4\z) circle (0.25cm);

            \draw[->] (root\z) -- (child1\z);
            \draw[->] (root\z) -- (child2\z);
            \draw[->] (child2\z) -- (leaf3\z);
            \draw[->] (child2\z) -- (leaf4\z);
        }
        
        \fill[orange, opacity=0.15, rounded corners] 
            (child28.north west) -- (child20.north east) -- (leaf40.south east) -- (leaf48.south west) -- cycle;
        \fill[magenta, opacity=0.15, rounded corners] 
            (child28.north west) -- (child20.north east) -- (leaf30.south east) -- (leaf38.south west) -- cycle;
        \fill[green, opacity=0.15, rounded corners] 
            (root8.north west) -- (root0.north east) -- (child10.south east) -- (child18.south west) -- cycle;
        \fill[blue, opacity=0.15, rounded corners] 
            (root8.north west) -- (root0.north east) -- (child20.south east) -- (child28.south west) -- cycle;

        \begin{scope}[xshift=0cm, yshift=5cm]
            \definecolor{nodecolor}{rgb}{0.55,0.6,0.75}
            \node[draw, rectangle, rounded corners, line width=1pt, inner sep=15pt, fill=green!15] (zoombox) at (0,0) {
                \begin{tikzpicture}
                    \foreach \i in {1,...,5}{
                        \path (\i*0.9, 1) coordinate (top\i);
                        \draw[fill=nodecolor, opacity=0.6] (top\i) circle (0.24cm);
                    }
                    \foreach \j in {1,...,5}{
                        \path (\j*0.9, -1) coordinate (bottom\j);
                        \draw[fill=nodecolor, opacity=0.6] (bottom\j) circle (0.24cm);
                    }
                    \foreach \i in {1,...,5}{
                        \foreach \j in {1,...,5}{
                            \draw[-, red, opacity=0.5, thin] (top\i) -- (bottom\j);
                        }
                    }
                \end{tikzpicture}
            };
        \end{scope}
        
        \draw[red, line width=1.5pt, rounded corners] (child10.north west) -- (-2.55, 3.25);
        \draw[red, line width=1.5pt, rounded corners] (root8.north west) -- (2.55, 6.75);
        
    \end{tikzpicture}
    }
\end{subfigure}

\caption{Directed graphical models for site-dependent mutation.
(a) Time-discretized CTMC with nearest-neighbor dependent rates. At each time 
$t_j$ nucleotide $x_i(t_j)$ is conditionally dependent only on local sequence context $(x_{i-1}(t_{j-1}),x_i(t_{j-1}),x_{i+1}(t_{j-1}))$ at 
$t_{j-1}$, 
but this local dependence propagates recursively over time yielding long-range dependence 
after marginalizing intermediate sequences. 
In continuous time, 
each 
$x_i(t)$ depends on the entire sequence $\mathbf{x}(0)$ at every $t$. (b,c) Phylogenetic trees under independent- vs dependent-site mutation.
Under 
independence (b) the model factors, 
and pruning can be performed at each site independently, but (c) under site dependence, nodes at each level become interdependent across all sites as in (a).}
\label{fig:graphical_model}
\label{fig:graphical_ism}
\label{fig:graphical_dsm_zoomed}
\end{figure}

Under an ISM, the marginal likelihood \eqref{eq:MargLik} of the tree $g$ factors site by site, as shown in Figure~\ref{fig:graphical_ism}b:
\begin{align}
p_0(Y \mid g, \Psi_0) 
= \prod\nolimits_{k=1}^\ell \sum\nolimits_{x_k, Z_k} p( Y_k, Z_k, x_k \mid g, \Psi_0)
\label{eq:MargLikISM}
\end{align}
%
with the inner sum 
over 
nucleotides 
at site $k$ of each unobserved sequence
, and 
can be computed efficiently 
Felsenstein 
pruning~\cite{felsenstein_evolutionary_1981}, 
but computing
(\ref{eq:MargLik}) 
is intractable for 
DSMs 
where 
this factorization 
fails (Figure~\ref{fig:graphical_dsm_zoomed}c).

%


The form \eqref{eq:MargLik} suggests a 
direct extension of the data augmentation importance sampler in
\cite{Mathews:2025a} for approximating pairwise sequence transition probabilities under DSMs, to phylogenetic trees.  
Specifically, after 
sampling trees from the posterior distribution over genealogies under the independent site model:
\begin{align}
g^{(j)} 
\, \sim \, \pi_0(g \mid Y, \Psi_0) 
\, \propto \, p_0(Y \mid g, \Psi_0) f_G( g) \qquad\quad  j=1,\ldots, m
\label{eq:ISMTreeSamples}
\end{align}
%
we wish to compute the \textit{importance weights} \(w( g^{(j)})\) under the DSM, defined by
\begin{equation}
w(g) \defeq 
\frac{\pi_1(g \mid Y, \Psi_1)}{\pi_0(g \mid Y, \Psi_0)}
\propto \frac{p_1(Y \mid g, \Psi_1)}{p_0(Y \mid g, \Psi_0)} 
=\sum\nolimits_{\x} \frac{p_1 ( Y \mid \x, g, \Psi_1) p_1 ( \x \mid g, \Psi_1 )}{p_0 ( Y \mid \x, g, \Psi_0 ) p_0 ( \x \mid g, \Psi_0 )},
\label{eq:TreeWeight}
\end{equation}
%
in order to approximate posterior expectations under $\pi_1$ by 
$\hat{\E}_{\pi_1}\big(h(g)\big) 
:= \sum_{i=1}^m 
\tilde{w}(g^{(j)})h(g^{(j)})
$ for normalized weights $\tilde{w}$.
%
%
Here $p_i(Y \mid g^{(j)}, \Psi_i)$ is the marginal likelihood of 
observed sequences $Y$ for 
sampled 
tree $g^{(j)}$ under model $i=0$ (ISM) or 1 (DSM).
While 
factorization (\ref{eq:MargLikISM}) enables efficient calculation of $p_0(Y \mid g, \Psi_0)$, exact 
computation of $p_1 ( Y \mid g, \Psi_1)$ is intractable. We instead approximate \eqref{eq:TreeWeight} itself by Monte Carlo integration. To simplify notation, 
we suppress 
conditioning on $\Psi$, and use $p_0$ and $p_1$ to distinguish the ISM and DSM.

A key to our approach is the factorization of the \textit{complete data likelihood} into edge transition probabilities when $(\x,Z)$ are observed: 
%
$p_i(Y,Z,\x\mid g) = p_i(\x \mid g) \prod_{(\bu, \bv, t) \in E 
} 
p_i(\bv \mid \bu,t)$,
%
as seen in \eqref{eq:MargLikEdgeProduct}.
Hence 
we can approximate \eqref{eq:TreeWeight} by 
sampling root and 
internal node sequences $\big(\x^{(i)}, Z^{(i)}\big) \sim p_0( \x, Z \mid Y, g)$ from the ISM conditional 
distribution 
%
%
via Felsenstein
pruning 
\cite{felsenstein_evolutionary_1981}
and forming the estimator
\begin{equation}
\hat{w}(g) = \frac{1}{m} 
\sum\nolimits_{j=1}^m \frac{p_1(\x^{(j)} \mid g)}{p_0(\x^{(j)} \mid g)} \prod\nolimits_{e \in E \left( Y, g, \x^{(j)}, Z^{(j)} \right)} \hat{w}(e).
\label{eq:TreeWeightEst}
\end{equation}
where $\hat{w}$ is an unbiased estimator of the edge weight
%
$\E(\hat{w}(e)) = w (e)  
\defeq \frac{p_1( \bv \mid \bu, t)}{p_0(\bv \mid \bu, t)}$.
%
\begin{figure}[htbp]
\centering
\begin{subfigure}{.74\textwidth}
\resizebox{.8\textwidth}{!}{
    \begin{tikzpicture}[scale=1,
        thick,
        node style flat/.style={
            circle,
            draw=cyan!60,
            fill=cyan!20,
            text=black,
            font=\scriptsize,
            minimum size=6mm,
            inner sep=0pt,
            align=center
        },
        node style flat nonleaf/.style={
            circle,
            draw=cyan!60,
            line width=1.5pt,
            dashed,
            fill=cyan!10,
            text=black,
            font=\scriptsize,
            minimum size=6mm,
            inner sep=0pt,
            align=center
        },
        node style path/.style={
            shape=ellipse,
            draw=blue!60,
            fill=blue!10,
            dashed,
            line width=1.5pt,
            inner sep=0pt,
            minimum height=6mm 
        }
    ]

    \def\ystep{0.55}
    \def\zstep{1.6}

    \begin{scope}[xshift=-5cm]
        \coordinate (S1) at (7*\ystep, 3*\zstep);
        \coordinate (A1) at (3*\ystep, 2*\zstep);
        \coordinate (B1) at (11*\ystep, 1.5*\zstep);
        \coordinate (C1) at (1*\ystep, 1*\zstep);
        \coordinate (D1) at (5*\ystep, 1*\zstep);
        \coordinate (E1) at (9*\ystep, 0*\zstep);
        \coordinate (F1) at (13*\ystep, 0*\zstep);
        \coordinate (G1) at (0*\ystep, 0*\zstep);
        \coordinate (H1) at (2*\ystep, 0*\zstep);
        \coordinate (I1) at (4*\ystep, 0*\zstep);
        \coordinate (J1) at (6*\ystep, 0*\zstep);
        \draw[dashed] (S1) -- (A1);
        \draw[dashed] (S1) -- (B1);
        \draw[dashed] (A1) -- (C1);
        \draw[dashed] (A1) -- (D1);
        \draw[dashed] (B1) -- (E1);
        \draw[dashed] (B1) -- (F1);
        \draw[dashed] (C1) -- (G1);
        \draw[dashed] (C1) -- (H1);
        \draw[dashed] (D1) -- (I1);
        \draw[dashed] (D1) -- (J1);
        \node[node style flat nonleaf] at (S1) {$\x$};
        \node[node style flat nonleaf] at (A1) {$\z^1$};
        \node[node style flat nonleaf] at (B1) {$\z^4$};
        \node[node style flat nonleaf] at (C1) {$\z^2$};
        \node[node style flat nonleaf] at (D1) {$\z^3$};
        \node[node style flat] at (E1) {$\y^5$};
        \node[node style flat] at (F1) {$\y^6$};
        \node[node style flat] at (G1) {$\y^1$};
        \node[node style flat] at (H1) {$\y^2$};
        \node[node style flat] at (I1) {$\y^3$};
        \node[node style flat] at (J1) {$\y^4$};
    \end{scope}

    \begin{scope}[xshift=3.5cm]
        \coordinate (S2) at (7*\ystep, 3*\zstep);
        \coordinate (A2) at (3*\ystep, 2*\zstep);
        \coordinate (B2) at (11*\ystep, 1.5*\zstep);
        \coordinate (C2) at (1*\ystep, 1*\zstep);
        \coordinate (D2) at (5*\ystep, 1*\zstep);
        \coordinate (E2) at (9*\ystep, 0*\zstep);
        \coordinate (F2) at (13*\ystep, 0*\zstep);
        \coordinate (G2) at (0*\ystep, 0*\zstep);
        \coordinate (H2) at (2*\ystep, 0*\zstep);
        \coordinate (I2) at (4*\ystep, 0*\zstep);
        \coordinate (J2) at (6*\ystep, 0*\zstep);

        \path let \p1=($(A2)-(S2)$) in (S2) -- (A2) node[midway, sloped, node style path, minimum width=9mm](p_S2_A2){} node[midway, sloped] {${\mathcal{P}}^{\mathbf{x} \z^1}$};
        \path let \p1=($(B2)-(S2)$) in (S2) -- (B2) node[midway, sloped, node style path, minimum width=9mm](p_S2_B2){} node[midway, sloped] {${\mathcal{P}}^{\mathbf{x} \z^4}$};
        \path let \p1=($(C2)-(A2)$) in (A2) -- (C2) node[midway, sloped, node style path, minimum width=9mm](p_A2_C2){} node[midway, sloped] {${\mathcal{P}}^{\z^1 \z^2}$};
        \path let \p1=($(D2)-(A2)$) in (A2) -- (D2) node[midway, sloped, node style path, minimum width=9mm](p_A2_D2){} node[midway, sloped] {${\mathcal{P}}^{\z^1 \z^3}$};
        \path let \p1=($(E2)-(B2)$) in (B2) -- (E2) node[midway, sloped, node style path, minimum width=9mm](p_B2_E2){} node[midway, sloped] {${\mathcal{P}}^{\z^4 \mathbf{y}^5}$};
        \path let \p1=($(F2)-(B2)$) in (B2) -- (F2) node[midway, sloped, node style path, minimum width=9mm](p_B2_F2){} node[midway, sloped] {${\mathcal{P}}^{\z^4 \mathbf{y}^6}$};
        \path let \p1=($(G2)-(C2)$) in (C2) -- (G2) node[midway, sloped, node style path, minimum width=9mm](p_C2_G2){} node[midway, sloped] {${\mathcal{P}}^{\z^2 \mathbf{y}^1}$};
        \path let \p1=($(H2)-(C2)$) in (C2) -- (H2) node[midway, sloped, node style path, minimum width=9mm](p_C2_H2){} node[midway, sloped] {${\mathcal{P}}^{\z^2 \mathbf{y}^2}$};
        \path let \p1=($(I2)-(D2)$) in (D2) -- (I2) node[midway, sloped, node style path, minimum width=9mm](p_D2_I2){} node[midway, sloped] {${\mathcal{P}}^{\z^3 \mathbf{y}^3}$};
        \path let \p1=($(J2)-(D2)$) in (D2) -- (J2) node[midway, sloped, node style path, minimum width=9mm](p_D2_J2){} node[midway, sloped] {${\mathcal{P}}^{\z^3 \mathbf{y}^4}$};
        
        \draw[solid] (S2) -- (p_S2_A2) -- (A2);
        \draw[solid] (S2) -- (p_S2_B2) -- (B2);
        \draw[solid] (A2) -- (p_A2_C2) -- (C2);
        \draw[solid] (A2) -- (p_A2_D2) -- (D2);
        \draw[solid] (B2) -- (p_B2_E2) -- (E2);
        \draw[solid] (B2) -- (p_B2_F2) -- (F2);
        \draw[solid] (C2) -- (p_C2_G2) -- (G2);
        \draw[solid] (C2) -- (p_C2_H2) -- (H2);
        \draw[solid] (D2) -- (p_D2_I2) -- (I2);
        \draw[solid] (D2) -- (p_D2_J2) -- (J2);

        \node[node style flat] at (S2) {$\x$};
        \node[node style flat] at (A2) {$\z^1$};
        \node[node style flat] at (B2) {$\z^4$};
        \node[node style flat] at (C2) {$\z^2$};
        \node[node style flat] at (D2) {$\z^3$};
        \node[node style flat] at (E2) {$\y^5$};
        \node[node style flat] at (F2) {$\y^6$};
        \node[node style flat] at (G2) {$\y^1$};
        \node[node style flat] at (H2) {$\y^2$};
        \node[node style flat] at (I2) {$\y^3$};
        \node[node style flat] at (J2) {$\y^4$};
    \end{scope}
    \end{tikzpicture}
}
\end{subfigure}
\hfill
\begin{subfigure}{.24\textwidth}
\resizebox{.55\linewidth}{!}{
\begin{tikzpicture}[node distance = 15mm and 15mm,
V/.style = {circle, draw, fill=gray!30}, every edge quotes/.style = {auto, font=\footnotesize, sloped}]
\begin{scope}[nodes=V]
    \node (x) at (0, 0) {\( \mathbf{x} \)};
    \node (s) at (0, -1.2) {\( \mathbf{s} \)};
    \node (y) at (-1, -2.2) {\( \mathbf{y} \)};
    \node (z) at (1, -2.2) {\( \mathbf{z} \)};
\end{scope}

\draw[->, thick] (x) edge[""] (s)
    (s) edge[""] (y)
    (s) edge[""] (z);
\end{tikzpicture}
}
\end{subfigure}
\caption{(a,b) Schematic of two-stage (stratified) sampling procedure. (a) \textbf{Stage 1:} Latent internal node sequences $(\x,\z^{1:4})$
are sampled
conditional on observed sequences $(\y^{1:6})$. Each sampled configuration defines a stratum. (b) \textbf{Stage 2:}  Conditional on each configuration 
from Stage 1, the marginal transition probability for each edge (e.g. $p_1 ( \z^1 \mid \x, t_1)$) 
is approximated by endpoint-conditioned importance sampling of path. Given 
internal sequences, these transition probabilities are mutually independent. (c) A \textit{star path} example.}
\label{fig:stratified-sampling}
\label{Fig:StarPathExample}
\end{figure}
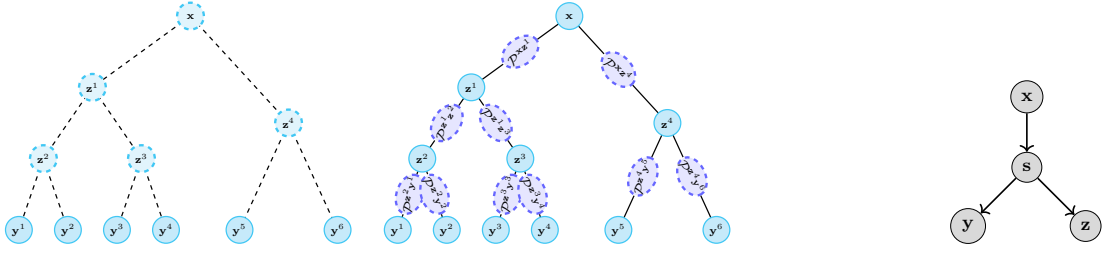
%
%

\noindent \textit{Estimating edge weights:} 
Computing the ISM transition probability $p_0(\bv\mid \bu,t)$ \eqref{eqn:ISM} for 
edge $e = ( \bu, \bv, t)$ is straightforward, 
but exact computation of $p_1(\mathbf{v} \mid \mathbf{u}, t)$ \eqref{Eqn:DSM}
remains intractable. \cite{Mathews:2024,Mathews:2025b} give 
an unbiased 
estimator $\hat{w}(e)$ of $w(e)$ with bounded approximation error, using SMC.
The SMC estimator uses 
a Gibbs sampler for 
sequence paths, which 
updates
the 
path at 
each 
site conditional on the 
paths of its neighboring sites based on the context-dependent model (DSM). This 
SMC-based unbiased estimator 
takes the form
\begin{equation} 
\label{eq:smc}
\hat{w}(e) = \prod\nolimits_{s = 1}^{S} \Big( \frac{1}{n} \sum\nolimits_{k = 1}^n w_s \big(\calP^{\left( k \right)}_s \left( e \right)\big) \Big) 
\defeq \prod\nolimits_{s=1}^S \hat{w}_s(e) \defeq \frac{\hat{p}_1(\bv \mid \bu,t)}{p_0(\bv \mid \bu,t)},
\end{equation}
where $\calP^{(k)}(e)$ is the $k$
th particle (evolutionary path from $\bu$ to $\bv$ in time $t$) for edge $e = (\bu, \bv, t)$, $w_s$ 
the corresponding weight at the $s$
th step, and $n$, $S$ 
the numbers of particles and 
steps, respectively.

\noindent \textit{Estimating tree weights:}
Plugging (\ref{eq:smc}) into (\ref{eq:TreeWeightEst}), we obtain an estimator for $w(g)$:
\begin{equation} 
\label{eq:TreeWeightEst2}
\hat{w}(g) = \frac{1}{m} \sum\nolimits_{j=1}^m \Big( \frac{p_1(\x^{(j)} \mid g)}{p_0(\x^{(j)} \mid g)} \prod\nolimits_{e \in E \left( Y, g, \x^{(j)}, Z^{(j)} \right)} \prod\nolimits_{s = 1}^S \hat{w}_s(e) \Big)
\defeq \frac{\hat{p}_1(Y, Z, \x \mid g) }{p_0(Y, Z, \x \mid g)}
\end{equation}
for $\x^{(j)}$ and $Z^{(j)}$ sampled from $p_0(\x, Z \mid Y, g)$, their joint conditional posterior distribution under the ISM.
%
%
%
%
The samples \eqref{eq:ISMTreeSamples} from $\pi_0( g \mid Y)$ are readily available from standard Bayesian phylogenetics software packages \cite{suchard_bayesian_2018}; 
our approach enables these to be extended to models 
with
site-dependence.
%
\textbf{Validation:}
Table~\ref{tab:validation} shows 
this approach on an example small enough (6 sequences of length $\ell=6$, 
Figure~\ref{fig:6SeqLen6}) 
for exact calculation via \eqref{Eqn:DSM}.
Posterior probabilities of all trees are accurately approximated with a relatively small sample 
(2000); however, 
the effective sample size (ESS) varies significantly across trees. Trees with lower ESS values (e.g. \#5) exhibit slightly larger errors, but not large enough to affect tree rankings. 

\begin{figure}[]
\begin{minipage}{0.22\textwidth}
\centering
\resizebox{.7\textwidth}{!}{
\begin{tikzpicture}[node distance = 15mm and 15mm,
V/.style = {circle, draw, fill=gray!30},
            every edge quotes/.style = {auto, font=\footnotesize, sloped}]
            \begin{scope}[nodes=V]
                \node (r) at (0, 0) {};
                \node (i1) at (-0.25, -0.5) {};
                \node (i2) at (0, -1) {};
                \node (i3) at (0.25, -1.5) {};
                \node (i4) at (0.5, -2) {};
                \node (t1) at (-1.25, -2.5) {};
                \node (t2) at (-0.75, -2.5) {};
                \node (t3) at (-0.25, -2.5) {};
                \node (t4) at (0.25, -2.5) {};
                \node (t5) at (0.75, -2.5) {};
                \node (t6) at (1.25, -2.5) {};
            \end{scope}
    
            \draw[->, thick] (r) edge[""] (i1)
                             (i1) edge[""] (t1)
                             (i1) edge[""] (i2)
                             (i2) edge[""] (t2)
                             (i2) edge[""] (i3)
                             (i3) edge[""] (t3)
                             (i3) edge[""] (i4)
                             (i4) edge[""] (t4)
                             (i4) edge[""] (t5)
                             (r) edge[""] (t6);
        \end{tikzpicture}
        }
        \caption{True tree}
        \label{fig:6SeqLen6}
    \end{minipage}
    \hfill
    \begin{minipage}{0.77\textwidth}
        \centering
        \captionof{table}{Exact vs approximated posterior prob. 
        for 6 sequence example in Fig.~\ref{fig:6SeqLen6}}
        \label{tab:validation}
        \begin{tabular}{|c|c|c|r||c|c|c|r|}
            \hline
            Tree & Exact Value & IS & ESS \; & Tree & Exact Value & IS & ESS \; \\
            \hline
            \( \#1 \) & \( -32.837 \) & \( -32.835 \) & \( 258.40 \) & 
            \( \#6 \) & \( -36.074 \) & \( -36.025 \) & \( 63.57 \) \\
            \hline
            \( \#2 \) & \( -34.193 \) & \( -34.129 \) & \( 125.05 \) &
            \( \#7 \) & \( -36.123 \) & \( -36.092 \) & \( 386.44 \) \\
            \hline
            \( \#3 \) & \( -34.642 \) & \( -34.672 \) & \( 44.32 \) &
            \( \#8 \) & \( -36.442 \) & \( -36.262 \) & \( 210.98 \) \\
            \hline
            \( \#4 \) & \( -34.869 \) & \( -34.905 \) & \( 233.28 \) &
            \( \#9 \) & \( -37.462 \) & \( -37.335 \) & \( 285.60 \) \\
            \hline
            \( \#5 \) & \( -35.121 \) & \( -35.186 \) & \( 36.40 \) &
            \( \#10 \) & \( -38.227 \) & \( -38.256 \) & \( 116.26 \) \\
            \hline
        \end{tabular}
        \end{minipage}
\end{figure}

\subsection{Sequential Monte Carlo} 
\label{sec:SMC}

To address the low ESS observed for some trees, we 
consider
an alternative 
approximation of \eqref{eq:MargLik},
by extending the SMC method of 
\cite{Mathews:2025b} from \textit{sequence paths} to \textit{tree paths}  
by introducing 
an MCMC sampler 
for 
the space $\mathscr{P}^{(Y, g)}$ of tree paths. The resulting method has 
similarities 
to the \textit{stepping stone (StSt)} method 
\cite{xie_improving_2010}; but 
whereas the StSt method approximates 
likelihood ratios 
for alternate ISM substitution models
 using 
 thermodynamic integration (TI)~\cite{gelman_simulating_1998,lartillot_computing_2006}, our approach 
 evaluates DSM likelihoods 
 using SMC 
 via importance sampling identities, 
 taking 
 advantage of 
 \textit{exact} computability of
 $p_0(Y \mid g, \Psi_0)$ \eqref{eq:MargLikISM} 
 given 
 $\Psi_0$.  For notational simplicity, 
 we assume the DSM is a \textit{nearest-neighbor} model 
 $\tilde{x_i} = (x_{i-1}, x_i, x_{i+1})$, but the 
 method 
 is general.

For a DSM \eqref{Eqn:CDrates}, we define a 
sequence of intermediate 
models 
%
$    \Tilde{\gamma}_t ( b ; \Tilde{x} ) = \phi^t ( b; \Tilde{x} ) \gamma ( b ; x )$
%
on a 
temperature ladder  
$\mathcal{T} = \{ 0 = t_0 < t_1 < \cdots < t_V = 1 \}$.
The SMC 
algorithm proceeds by: 
\begin{enumerate}
\item \textit{(Initialization)} Sample 
tree-path particles 
$\bPa^{(Y, g)}_{0,1}, \ldots, \bPa^{(Y, g)}_{0,N} \stackrel{\text{iid}}\sim \pi_0 (\cdot \mid Y,g)$ 
under the ISM, performed 
site-by-site by 
(i) sampling internal node sequences using Felsenstein
pruning 
and then (2) 
sampling endpoint-conditioned substitution paths along each branch under the ISM~\cite{hobolth_simulation_2009,rodrigue_uniformization_2008}.
\item For $v = 1, \ldots, V$:
\begin{enumerate}[label=(\roman*)] 
    \item \textit{(Resampling)} Sample $\Tilde{\bPa}^{(Y, g)}_{v, 1}, \ldots, \Tilde{\bPa}^{(Y, g)}_{v, N}$ with probability 
 $\propto w_v (\bPa^{(Y, g)}_{v-1, i})
  := \frac{p_{t_v}(Y,\bPa^{(Y, g)}_{v-1, i} \mid g)}{p_{t_{v-1}}(Y,\bPa^{(Y, g)}_{v-1, i} \mid g)}$.
    %
    \item \textit{(Mutation)} Sample ${\bPa}_{v, 1}^{(Y, g)}, \ldots, {\bPa}_{v, N}^{(Y, g)}$ where ${\bPa}_{v, i}^{(Y, g)} \mid \Tilde{\bPa}_{v, i}^{(Y, g)} \sim \mathbf{K}_{v}^{s} ( \Tilde{\bPa}_{v, i}^{(Y, g)}, \cdot )$ and \( \mathbf{K}_{v} \) is an ergodic \( \pi_{t_{v}} \)-invariant Markov kernel.
    \end{enumerate}
\item Form the estimate
%
$\hat{p}_1 (Y \mid g) = p_0 (Y \mid g) \prod_{v=1}^V \frac{1}{N} \sum_{i=1}^N w_v(\bPa^{(Y, g)}_{v-1, i}). \hfill (14)$
%
\end{enumerate}
Step 2(i) describes multinomial resampling for simplicity of presentation; in practice alternative variance-reduction techniques are preferred for 
resampling 
\cite{Doucet:2001}.  Step 2(ii) is implemented by simulating the evolution of each particle according to the Markov kernel $K_v$ for $s$ steps. We therefore require a $\pi_t$-invariant Markov kernel, which we now construct. Specifically, we extend the MCMC kernel of~\cite{li_on_2025} for pairwise endpoint-conditioned sequence paths to a MCMC kernel on tree-paths. The pairwise kernel is a Metropolis–Hastings chain using the Hobolth algorithm~\cite{hobolth_markov_2008} as a proposal distribution (see Appendix for additional details).

\textbf{Star-path MCMC kernel:} We introduce
a Markov kernel on tree-path space that updates tree-paths component-wise by (randomly or systematically) selecting a tree node $\s$ and proposing  a \textit{star-path} change; see Figure~\ref{Fig:StarPathExample} (each of the four nodes represents a sequence of length $\ell$).
Let $\s_{-i} = \{ s_1, \ldots, s_{i-1}, s_{i+1}, \ldots, s_{\ell} \}$, and let $\bPa^{(\x\s)} := (\Pa^{(\x\s)}_1, \ldots, \Pa^{(\x\s)}_{\ell} )$ 
be an evolutionary path from $\x$ to $\s$
with $\Pa^{(\x\s)}_j$ 
the 
path at site $j$ from $x_j$ to $s_j$. 
The kernel draws a \textit{single-site star-path} $\bPa^{(\x\s\y\z)}_i := ( \Pa^{(\x\s)}_i, \Pa^{(\s\y)}_i, \Pa^{(\s\z)}_i)$ conditional on 
paths at all other sites, i.e. $p_1 (\bPa^{(\x\s'\y\z)}_i \mid \bPa^{(\x\s'\y\z)}_{[-i]})$. A detailed description of this MCMC procedure is given in Appendix~\ref{Sec:AppdxStartPath}.

\subsection{Comparison}

Table~\ref{Tbl:ESSComparison} compares 
Stratified Sampling 
\eqref{eq:TreeWeightEst2}, 
Importance Sampling, and 
SMC (14)
for equal 
particle numbers, and for 
equal computational budget. The results show 
marked differences: SMC consistently surpasses SS and IS, achieving 
substantially higher 
ESS ($\geq 971.71$/step with 1024 particles) in both scenarios, two orders of magnitude greater than either SS or IS. Under equal computational budgets, the simplicity of IS allows 
higher throughput 
($2.56\times 10^6$ particles (ESS 137.33) to SMC's 1024), 
but SS manages only a $4\times$ increase and
little ESS improvement.
The surprisingly dominant performance of SMC can be explained by more efficient allocation of samples in 
$(\x,Z)$ space by the MCMC chain relative to the ISM proposal; see Appendix~\ref{sec:AppdxVar}.

\begin{table}[htbp]
\centering
    \begin{tabular}{|lrr||lrr|}
        \multicolumn{3}{c}{Fixed Particle Count} &
\multicolumn{3}{c}{Fixed Computational Budget}
\\
        \hline
        Method & Particles ($N$) & ESS & Method & Particles ($N$) & ESS \\ 
        \midrule
        SS & $1,024$ & $1.11$ &  SS & $4,096$ & $1.22$\\ 
        IS & $1,024$ & $4.88$ & IS & $2,560,000$ & $137.33$\\ 
        SMC & $1,024$ & \quad $\geq 971.71$ & SMC & $1,024$ & \quad $\geq 971.71$\\ 
     \bottomrule
    \end{tabular}
\caption{
Effective sample sizes for importance sampling (IS), stratified IS (SS), and sequential Monte Carlo (SMC) estimators of posterior probability under the \Arm model, 
for a phylogenetic tree of the DH270 clone sampled by \texttt{linearham}.
ESS for SMC is per step.  
Under fixed computational budget, 
all algorithms were run for the same CPU time required for SMC.}
\label{Tbl:ESSComparison}
\end{table}

\subsection{Incorporating VDJ Model-Based Priors on the UCA}
\label{Sec:VDJPrior}

Most 
applications of phylogenetic 
inference assume 
the common ancestor $\x$ is drawn from the stationary distribution of the evolutionary model. However, solving for the stationary distribution of a CTMC model with arbitrary site-dependence can 
be difficult, with 
the chain often non-reversible. (An exception is if 
the model satisfies conditions guaranteeing 
the stationary distribution 
is a Markov random field; see~\cite{jensen_probabilistic_2000}.) 
For reconstructing BCR 
lineages, this assumption is not appropriate: the UCA is generated 
by VDJ recombination
of germline sequences.
Stochastic models of 
VDJ recombination 
are well-established~\cite{elhanati_inferring_2015,ralph_consistency_2016}, and these models can provide informative prior distributions on the UCA during BCR lineage reconstruction~\cite{Ralph:2016,dhar_bayesian_2020}. 
%
Prior distributions on the UCA 
can be incorporated 
into the importance sampling approaches of Sections~\ref{sec:IS} and~\ref{sec:SMC}. 
Let $\zeta_0 (\x)$ denote the marginal distribution of $\x$ in the 
importance sampling proposal (e.g. the stationary distribution under 
ISM $p_0$), and 
$\zeta_1(\x)$ 
the desired prior,
then
we have $p_i(Y,\x,Z \mid g) = \zeta_{i} (\x) \prod_{(\mathbf{u}, \mathbf{v}, t) \in E \left( g, \x, Z, Y \right)} p_i(\mathbf{v} \mid \mathbf{u},t) $ and the importance weights can be calculated as before. However, 
for the VDJ model these priors may place 
mass on significantly different sequences, leading to 
inflation of importance weights. To prevent this, we instead we use the approach of \cite{dhar_bayesian_2020}  to sample from the posterior defined by the VDJ model combined with the ISM model for SHM:
%
$\pi_{0,\text{VDJ}} \propto p_0 (Y, Z \mid \x, g)\zeta_{\text{VDJ}}(\x)$.
%
In this case the prior $\zeta_{\text{VDJ}}$ cancels out of 
the importance weights, and the estimator~\eqref{eq:TreeWeightEst2}  remains unchanged.

\paragraph{Piecewise VDJ prior:}
In Section~\ref{Sec:Results}, we show that the VDJ prior improves reconstruction on germline-encoded regions, but sometimes causes 
errors in non-templated regions. We therefore define 
a piecewise prior under which the VDJ model applies outside the CDR3 region but \textit{not} within the CDR3 itself. To draw from the 
posterior 
we 
use \texttt{linearham} \cite{dhar_bayesian_2020} to sample 
$\pi_{0,\text{VDJ}}$ 
then substitute a CDR3 drawn from $p_0(\x_{\text{CDR3}} \mid Y,g)$ by pruning, resulting in the piecewise prior $p_{\text{VDJ}} ( \x_{\overline{\mathrm{CDR3}}} )\zeta_{0} ( \x_{\mathrm{CDR3}})
$, and the 
weight 
\eqref{eq:TreeWeight} remains correct.
%

\subsection{Bayes Estimators for the UCA}

Bayesian inference results in a posterior distribution $\pi_1(\x,g \mid Y)$, but 
users often prefer to obtain a single reconstructed lineage $g$ or UCA $\x$, requiring 
summarization of the posterior $\pi_1$ by a point estimate. To evaluate 
reconstruction 
accuracy 
we consider the \textit{maximum a posteriori} (MAP) estimator $\hat{\x}_{\text{MAP}}$ defined by the sampled UCA with 
highest estimated posterior probability and the \textit{marginal mode} (MM) estimator $\hat{\x}_{\text{MM}}$ which concatenates the nucleotides maximizing the marginal posteriors at each sequence position; explicit formulas for these estimators are given in Appendix~\ref{Sec:AppedxUCAEst}.

\section{Results}
\label{Sec:Results}

We first describe experiments on simulated datasets designed to demonstrate the effectiveness of our approach, and to investigate impact of accounting for context-dependence on the reconstruction of phylogenetic trees (B cell lineages) and ancestral sequences (UCAs).  We then apply our approach to the analysis of BCR repertoires for two important 
broadly-neutralizing HIV antibodies  DH270 and CH235.

In each case, inference is performed according to the importance sampling approaches of  Section~\ref{sec:methods}. Samples from the posterior distribution of trees under the pure phylogenetic ISM were generated using Beast~\cite{suchard_bayesian_2018}
under a Yule process tree prior and assuming the strict clock model.  Inference was performed using two MCMC chains with 10M iterations each, with a burn-in of 6M  and thinning of 1k. Convergence was diagnosed by the ESS (as reported by Beast) 
for both the joint probability under the ISM and the root height of trees, so the number of root and internal node sequence samples varies by problem.

\subsection{Simulation study design} 

 To evaluate the accuracy of our approach on data for which the ground truth (genealogy and UCA) are known, we simulated multiple sets of BCR sequences, each representing a clonal repertoire dataset. 
Sequences were simulated using a neutrally evolving branching process consisting of a \(\text{Pois}(\lambda) \) progeny distribution  and a \(\text{Pois}(\lambda_{\text{len}})\) branch length generating distribution \cite{davidsen_benchmarking_2018}, starting from a single naive sequence (the UCA).
The number of progeny cells is determined by a $\mathrm{Pois}( \lambda)$ draw: zero means the cell dies, one means no division, and two or more means the cell splits. Each progeny cell then undergoes evolution for a time interval drawn from $\mathrm{Pois}(\lambda_{\text{len}})$, with somatic hypermutations occurring sequentially and incorporating context-sensitivity. The simulation concludes when all cells die or a pre-determined time is reached~\cite{davidsen_benchmarking_2018}. In what follows, we set \( \lambda = 1.5 \) and \( \lambda_{\text{length}} = 0.005 \), similar to real SHM rates~\cite{davidsen_benchmarking_2018}.

\subsection{Effects of context-dependence on tree and UCA reconstruction accuracy}
\label{sec:ResultsVDJ}

\subsubsection{Short Sequences: CDR3 regions}
The complementarity-determining region 3 (CDR3) of antibodies
is a region of high variability that plays a crucial role in antigen recognition and binding. 
Therefore, we first focus on short sequences simulated from the CDR3 of a bnAb UCA.
Two
independent (tree, clonal sequence repertoire) sets were simulated.  
Samples from the posterior distribution over trees under the pure phylogenetic ISM model were obtained using Beast, and posterior 
probabilities under the DSM calculated from importance weights \eqref{eq:TreeWeightEst2}.
%
Results for UCA reconstruction are summarized in Table~\ref{Tbl:SSAccuracy}a.
The DSM consistently outperforms 
the ISM for all estimators.  Also shown are UCAs reconstructed conditional on the true tree (unknown in practice); unsurprisingly, this significantly reduces UCA reconstruction error, but moreso for the ISM. Thus a key role of the DSM appears to be in improving the estimated tree.  

\subsubsection{Full length BCR sequences}
Next we 
simulated 8 independent (tree, clonal sequence repertoire) sets of full length 
BCR sequences from an estimated UCA for HIV bnAb CH235.
Each tree was pruned to select a subset of nodes to match 
individual Hamming distances of a real BCR dataset while ensuring 
no selected node was an ancestor/descendant of another,  
to approximate fixed-depth sequencing of BCR repertoires. 
%
Reconstruction accuracy 
is shown in Table~\ref{Tbl:SSAccuracy}a. 
In each simulated 
dataset, the true tree ranks as the highest 
probability under the DSM. 
For these longer sequences the number of reconstruction errors is larger, and the improvement obtained under the DSM 
is even clearer. 
(In Table~\ref{Tbl:SSAccuracy} ISM refers to the mean-field DSM approximation (see Appdx~\ref{sec:AppExp}); SHM to the pure phylogenetic model; VDJ to Linearham. 
IgPhyml does not do CDR3 junction inference \cite{hoehn_repertoire-wide_2019} so does not provide a meaningful comparison.)


\subsubsection{Improving UCA reconstruction with VDJ recombination models:}
In standard phylogenetics, the common ancestor $\x$ is assumed to be drawn from the stationary  distribution of the evolutionary  model. In affinity maturation, the UCA
is generated by the mechanisms of VDJ recombination. \cite{hozumi_evidence_1976,tonegawa_somatic_1983}. 
Detailed stochastic models of this process have been developed \cite{dhar_bayesian_2020,kepler_reconstructing_2013,Marcou2018} that assign a probability to every possible UCA sequence arising from the host germline VDJ gene segment library via recombination and $n$-nucleotide insertion. Although developed for BCR sequence analysis including the specific purpose of lineage reconstruction and UCA estimation, these models have relied primarily on 
independent-site models of SHM. However, VDJ models can provide additional information regarding the UCA prior to SHM, and incorporating this information may be expected to improve reconstruction accuracy. Table~\ref{Tbl:SSAccuracy}a shows the results of incorporating the HMM-based VDJ rearrangement model of \cite{dhar_bayesian_2020}. As expected, the accuracy of UCA reconstruction does indeed significantly improve, although accuracy obtained by our importance sampling-based DSM estimator is constrained by the failure of the ISM-based \texttt{linearham} model to sample the correct UCA at all.
Notably, all observed reconstruction errors occur in the CDR3 region, suggesting that the prior information of the VDJ model significantly improves accuracy in the germline-templated V, and J regions, but is less accurate in the highly variable $n$-nucleotide inserted regions, where it may actually introduce errors.
%
\textbf{Piecewise VDJ prior:}
Table~\ref{Tbl:SSAccuracy}a shows that substituting the piecewise prior described in section~\ref{Sec:VDJPrior} for the full VDJ prior does indeed provide a small further improvement in UCA reconstruction accuracy.
\begin{table}[ht]
    \centering
    \resizebox{.4\textwidth}{!}{
    \begin{tabular}{|c|c|c|c|c|c|c|c|c|c|}
        \hline
         Dataset & \multicolumn{3}{|c|}{CDR3} & \multicolumn{6}{|c|}{Full Length BCR (SS)} \\
        \hline
        Model & \multicolumn{3}{|c|}{SHM} & \multicolumn{2}{|c|}{SHM} & \multicolumn{2}{|c|}{
        +VDJ} & \multicolumn{2}{|c|}{
        +piecewise} \\
        \hline
        CDR3 & \multicolumn{3}{|c|}{
        Yes} & 
        Yes & 
        No & 
        Yes & 
        No & 
        Yes & 
        No \\
        \hline
        Estimator & $\bar{T}$
        & $T_{\text{MAP}}$ 
        & $T^\star$ 
        & $\bar{T}$ & $\bar{T}$ &  $\bar{T}$ & $\bar{T}$ & $\bar{T}$ &  $\bar{T}$\\
        \hline
        \(\hat{\E}_{p_0}[d(\cdot,\x^\star)]\) 
         & \( 6.6 \) & \( 6.0 \) & \( 1.2 \) & \( 1.2 \) & \( 16.5 \) & \( 3.2 \) & \( 0 \) & \( 2.4 \) & \( 0 \) \\
        \hline
        \(\hat{\E}_{p_1}[d(\cdot,\x^\star)]\) & \( 4.9 \) & \( 5.3 \) & \( 1.0 \) & \( 1.0 \) & \( 13.9 \) & \( 3.0 \) & \( 0 \) & \( 1.9 \) & \( 0 \) \\
        \hline
        UCA$^{\text{MAP}}_{\text{ISM}}$ & \( 6 \) & \( 7 \) & \( 1 \) & \( 1 \) & \( 16 \) & \( 3 \) & \( 0 \) & \( 2 \) & \( 0 \) \\
        \hline
        UCA$^{\text{MM}}_{\text{ISM}}$ & \( 7 \) & \( 7 \) & \( 1 \) & \( 1 \) & \( 13 \) & \( 3 \) & \( 0 \) & \( 2 \) & \( 0 \) \\
        \hline
        UCA$^{\text{MAP}}_{\text{DSM}}$ & \( 5 \) & \( 5 \) & \( 1 \) & \( 1 \) & \( 12 \) & \( 2 \) & \( 0 \) & \( 1 \) & \( 0 \) \\
        \hline
        UCA$^{\text{MM}}_{\text{DSM}}$ & \( 5 \) & \( 5 \) & \( 1 \) & \( 1 \) & \( 11 \) & \( 2 \) & \( 0 \) & \( 1 \) & \( 0 \) \\
        \hline
    \end{tabular}
    } \qquad
    \resizebox{.45\textwidth}{!}{
    \begin{tabular}{|c||c|c|c|c|}
    \hline
    \multicolumn{5}{|c|}{Full Length BCR (SMC), with NGS-VDJ}\\
        \hline
         & 
         ISM & 
         DSM & mean diff. [range] & \% of UR\\
        \hline
        $\hat{\E}[d(\cdot,\x^{*})]$ & 
        \textbf{3.5} 
        & 
        \textbf{2.0} 
        & 1.45 \ [0.25, 2.63] & 
        4.8\%\\
        \hline
        $d(\hat{\x}_{\text{MM}}, \x^{*})$ & 
        \textbf{2.6}
        & 
        \textbf{2.2}
        & 0.40 \ [0, 2] & 
        1.3\%\\
        \hline
        $d(\hat{\x}_{\text{MAP}}, \x^{*})$ & 
        \textbf{2.8} 
        & 
        \textbf{2.2} 
        & 0.60 \ [0, 2] & 
        2.0\%\\
        \hline
        \hline
        & 
        ISM $\mid g^*$ & 
        DSM $\mid g^*$ & mean diff. [range] & \% of UR\\
         \hline
        $\hat{\E}[d(\cdot,\x^{*})]$ & 
         \textbf{2.4} 
        & 
        \textbf{2.0}
        & 0.39 \ [-0.20, 0.93] & 
        1.3\%\\
        \hline
        $d(\hat{\x}_{\text{MM}}, \x^{*})$ & 
        \textbf{2.4}
        & 
        \textbf{1.8}
        & 0.60 \ [0, 1] & 
        2.0\%\\
        \hline
        $d(\hat{\x}_{\text{MAP}},\x^{*})$ & 
        \textbf{2.4}
        & 
        \textbf{1.8}
        & 0.60 \ [0, 1] & 
        2.0\%\\
        \hline
    \end{tabular}
    }
       \caption{(a) Hamming distance between estimated and true UCA $\x^\star$, for various models and estimators described in text. $\bar{T}$ denotes marginalization over the posterior distribution on trees; $T_{\text{MAP}}$ and $T^\star$ denote conditioning on the MAP and true trees, respectively. Errors are divided into CDR3 and non-CDR3 regions. (b) Hamming distance using SMC algorithm and NGS-retrained VDJ model,
    averaged over 10 independent simulations. Differences show improvement of DSM over ISM in terms of sites and as \% of uncertain region (UR) consisting of N-nucleotides and D gene segments (average length: 30). Differences occur at a small fraction of sequence positions, but ISM errors are at functionally critical sites (Sec.~\ref{sec:HIVbnAbs}).}
\label{Tbl:SSAccuracy}
\label{Tbl:SMCFullSeqAccuracy}
\end{table}

\subsubsection{VDJ Reparameterization:}
Results in Table~\ref{Tbl:SSAccuracy}a incorporating the VDJ prior were obtained using default parameters for \texttt{linearham}.  To test the sensitivity of these results to the parameterization of the VDJ model, we re-ran simulation studies using a parameter set for \texttt{linearham} obtained by training on a large NGS dataset of human B cell repertoire heavy chain sequences. 
10 independent simulated clonal datasets were generated using UCAs sampled from the stochastic VDJ rearrangement process (via \texttt{linearham}) and a tree topology (including branch lengths) sampled from the Yule prior.
Observed sequences were then obtained by simulating context-dependent evolution  under the ARMADiLLO model along the sampled tree topology. These simulated repertoires match many key statistics of  real-world clones, with a typical clone having sequence length 
around 365 and containing 55 highly-mutated observed sequences,  
with average Hamming distance 
from the UCA of 84.44 ($\approx 75\%$  sequence identity),  
average tree depth 
similar to that 
of trees reconstructed for the CH235 clone, and depth of the leaf closest 
to the root (which provides most information about the UCA) of $0.08$ with 27.35 mutations from the root sequence on average (least mutated sequence in real CH235 clone dataset: 27 mutations). SMC was used in all reconstructions based on results in
Table~\ref{Tbl:ESSComparison}.

\begin{figure}[htbp]
\centering
\begin{subfigure}[b]{0.36\textwidth}
\centering
\includegraphics[width=.83\linewidth,
trim = .15in .2in .15in .5in,clip]{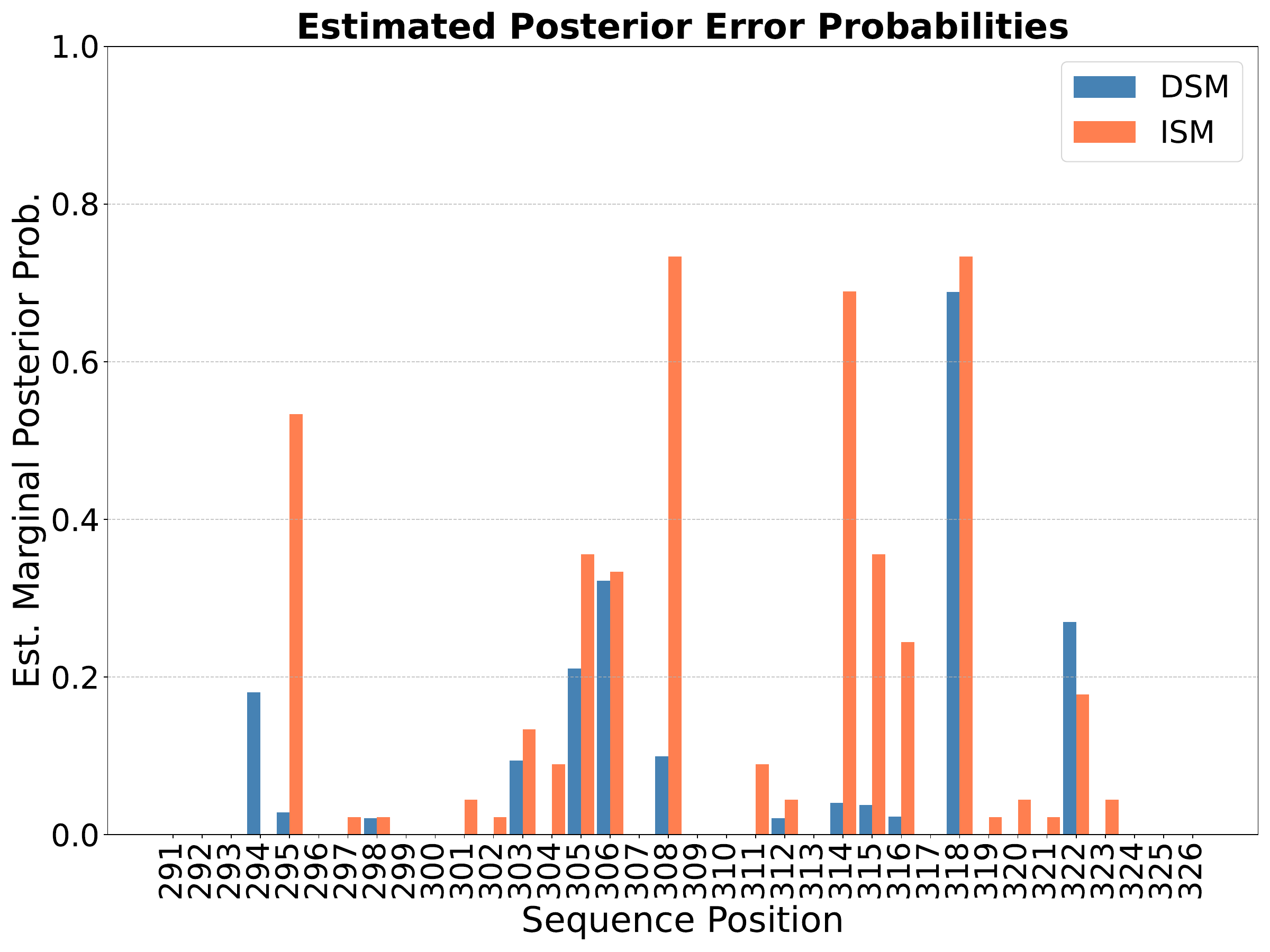}
\caption{Marginalized over trees.}
\end{subfigure}
\begin{subfigure}[b]{0.31\textwidth}
\centering
\includegraphics[width=.95\linewidth, 
trim = .5in .1in .5in .5in,clip]{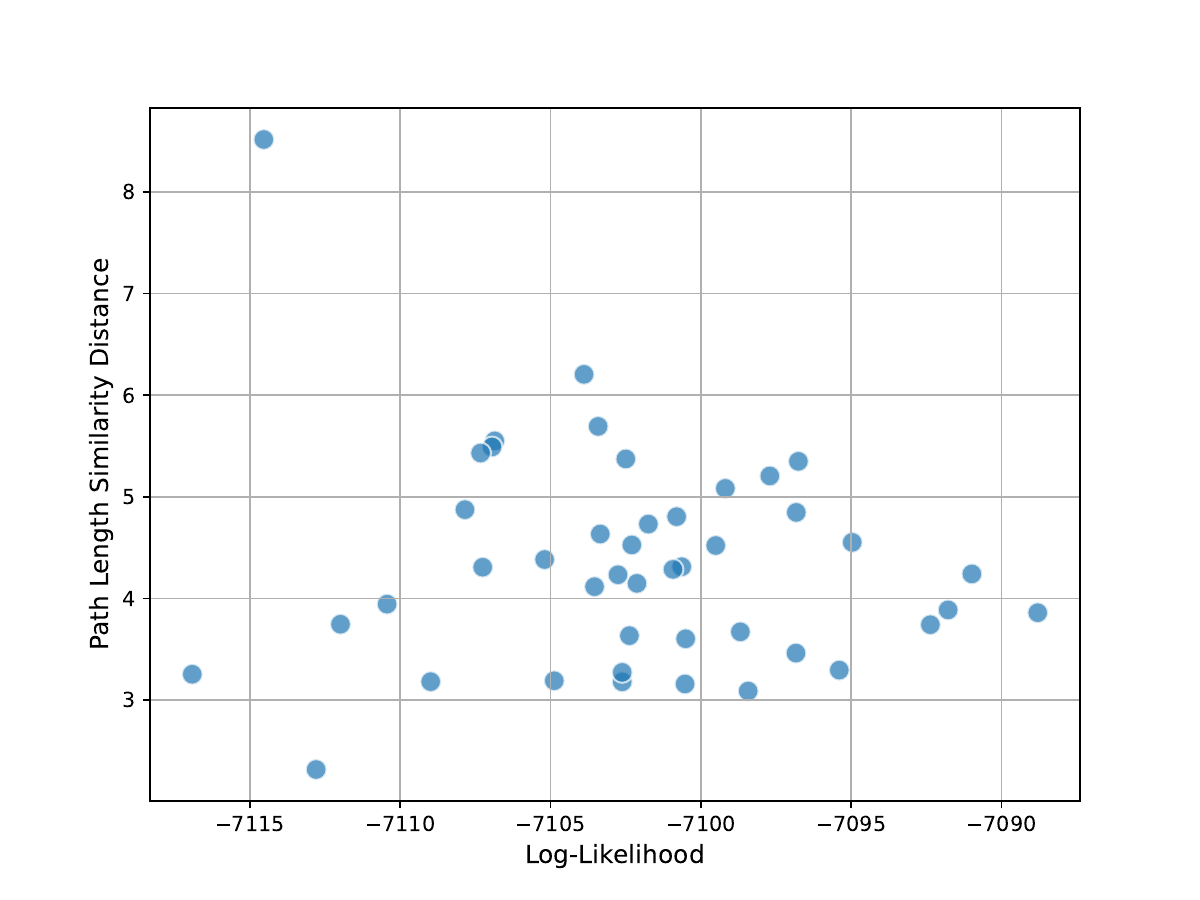}
\caption{correlation -0.145.}
\label{fig:ism_pls}
\end{subfigure}
\begin{subfigure}[b]{0.31\textwidth}
\centering
\includegraphics[width=.95\textwidth,
trim = .5in .1in .5in .5in,clip]{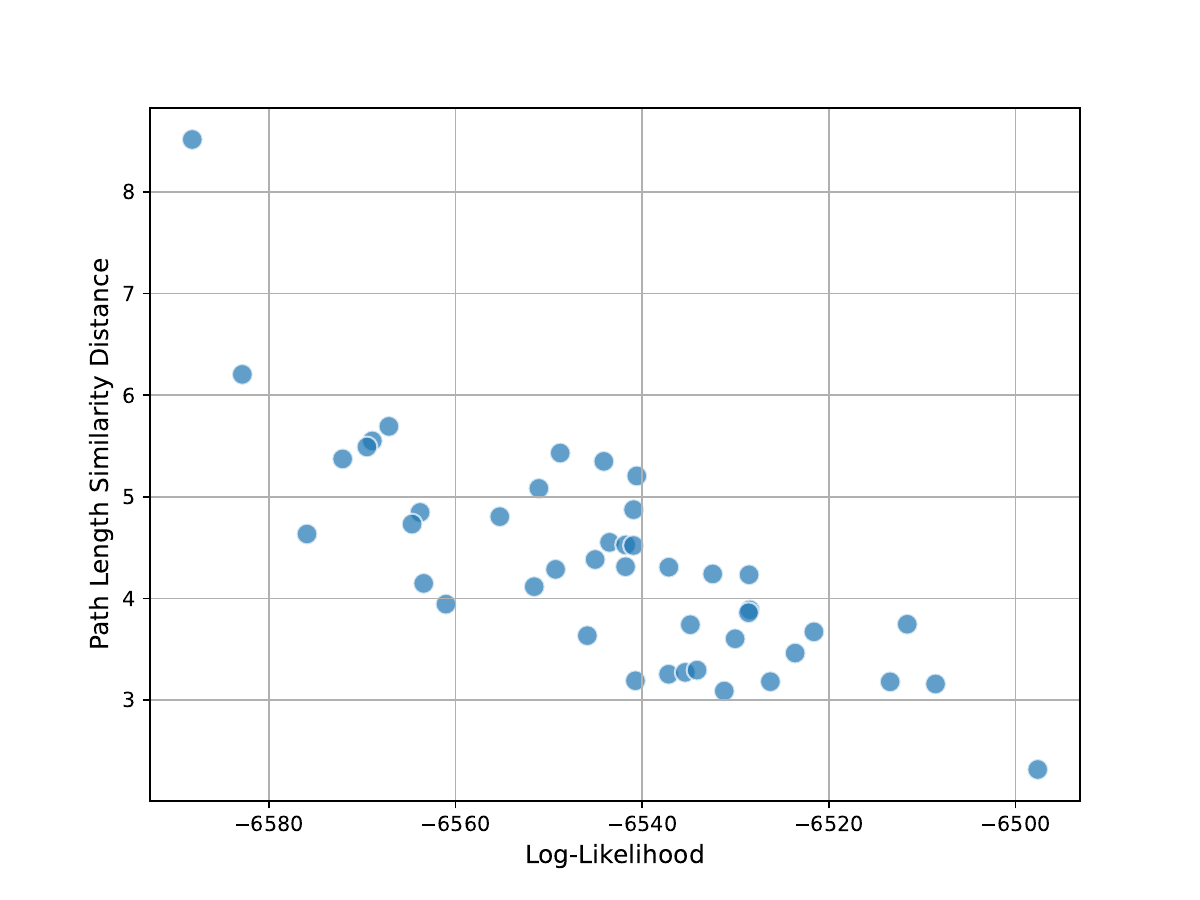}
\caption{correlation -0.789.}
\label{fig:dsm_pls}
\end{subfigure}
\caption{(a) 
Marginal probability of reconstruction error 
by UCA position under ISM vs DSM.
The DSM shows reduced errors at all positions 
except bases 
294 and 322. Most errors occur in positions 314, 315, and 316 that have
neighboring mutations which 
are more accurately reconstructed 
by accounting for context-dependence. (b,c) Tree 
likelihood 
vs. (path length similarity) distance 
from true tree, under (b) ISM and (c) DSM. Posterior mean distance 
is 3.89 (ISM) and 2.31 (DSM). 
The sampled tree with smallest plsd to the true tree ranks 3rd-to-last under the ISM among all sampled trees, but 1st (most probable) under the DSM.}
\label{fig:sim-err}
\end{figure}

Table~\ref{Tbl:SMCFullSeqAccuracy}b shows the resulting 
DSM  
error rate is 6.6\% versus 11.6\% for the ISM, a reduction of 43.1\%. The retrained VDJ prior no longer induces errors in the CDR3.  (However, most users will not have access to a large application-specific NGS dataset to retrain \texttt{linearham}, so we  still recommend the piecewise prior for robustness as in Table~\ref{Tbl:SMCFullSeqAccuracy}a.) Again 
a major source of error in UCA reconstruction comes from inaccuracies in the inferred tree; 
the DSM is able to more accurately reconstruct the tree, which in turn leads to improved accuracy in the UCA reconstruction.
Figure~\ref{fig:sim-err}a shows estimated marginal probability of UCA reconstruction errors by site for one simulated clone, demonstrating the significant improvement under the DSM.
%
%
Figure~\ref{fig:sim-err}b,c shows log-likelihood of trees versus distance to the true tree for both the ISM and DSM 
using 
path length similarity distance~\cite{williams_comparison_1971,steel_distributions_1993}, chosen due to 
vaccinologists' emphasis on the temporal trajectory of antibody lineages. Log-likelihoods under the DSM  correlate more strongly with proximity to the true tree.


\subsubsection{Observed ISM mistakes} 
\begin{figure}[ht]
\centering
\begin{subfigure}{0.24\textwidth}
\centering
\includegraphics[width=0.55\linewidth]{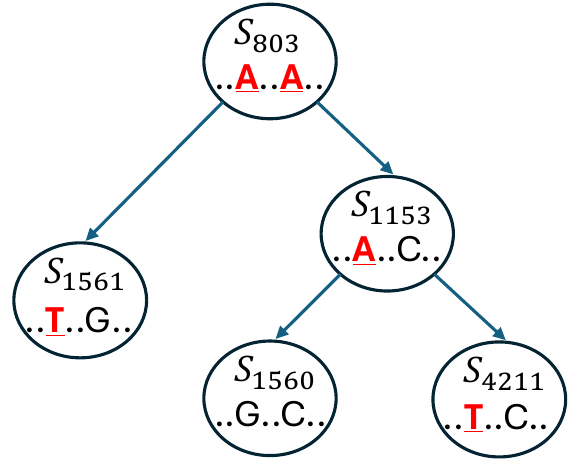}
\caption{True subtree}
\label{fig:scenario_1_true}
\end{subfigure}
\hfill
\begin{subfigure}{0.24\textwidth}
\centering
\includegraphics[width=0.55\linewidth]{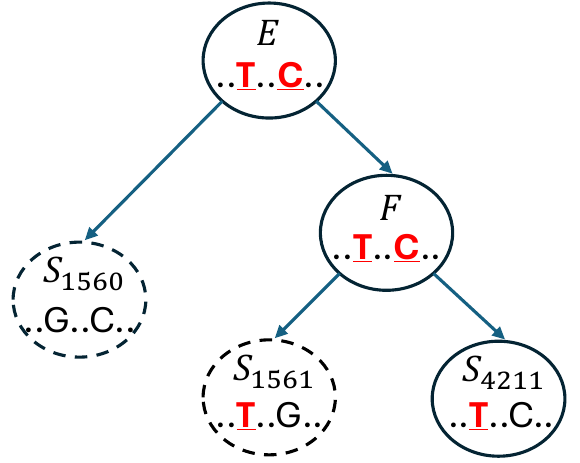}
\caption{ISM sampled subtree} 
\label{fig:scenario_1_sampled}
\end{subfigure}
\begin{subfigure}{0.25\textwidth}
\centering  \includegraphics[width=0.6\linewidth]{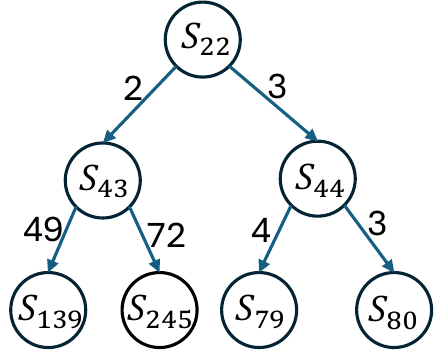}
\caption{True subtree}  \label{fig:scenario_3_true}
\end{subfigure}
\hfill
\begin{subfigure}{0.23\textwidth}
\centering        \includegraphics[width=0.55\linewidth]{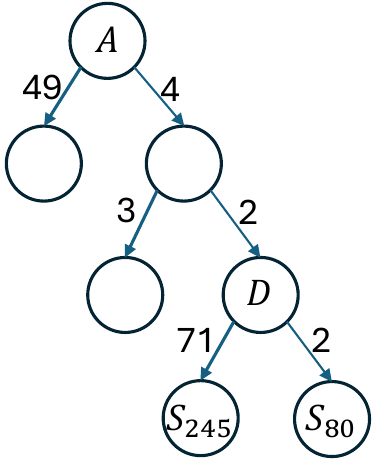}
\caption{ISM sampled subtree}        \label{fig:scenario_3_sampled}
\end{subfigure}
\caption{Errors in ISM tree reconstruction.  (a,b) The ISM essentially reconstructs the root sequence by finding common bases, but
highly mutable bases (red) in the UCA are less 
conserved at leaves, 
and finding 
common bases can 
fail.
(Dashed cirlces show mistakes in inferred tree.) 
(c,d) The 4 of the 8 mutations along $A\to D$ that are shared by $S_{245}$ and $S_{80}$ are highly mutable under the DSM.} 
\label{fig:scenarios}
\end{figure}

To understand the impact of the independent site assumption in the presence of site-dependence, we identified  mistakes commonly observed under 
ISM reconstruction. These illustrate how failure to account for context-dependent mutation leads to inaccuracies in phylogenetic tree reconstruction.
%
\paragraph{Scenario 1:} Figure~\ref{fig:scenario_1_true} shows a small subtree of a larger tree, with 
\( \mathrm{S}_{1560} \), \( \mathrm{S}_{1561} \) and \( \mathrm{S}_{4211} \) 
observed leaves. It thus represents a tree reconstruction problem with only 3 leaf nodes and
observed sequences of length 367. Despite the large amount of information,
the ISM-reconstructed topology 
is incorrect.
%
%
Under the ISM, the reconstructed tree minimizes the sum of Hamming distances along 
edges. 
Key sites are 147 and 188: 
Figure~\ref{fig:scenarios} shows that \( \mathrm{S}_{1561} \) and \( \mathrm{S}_{4211} \) agree on \( \mathrm{T} \) at site \( 146 \) while \( \mathrm{S}_{1560} \) and \( \mathrm{S}_{4211} \) agree on \( \mathrm{C} \) at site \( 188 \),
hence the ISM estimates the UCA as
\( \mathrm{T} \) at site \( 147 \) and \( \mathrm{C} \) at site \( 188 \) with high probability, ensuring 
each site has only experienced a single mutation. However the DSM assigns a lower 
probability to the ISM 
tree
since these two key mutations occur 
in highly mutable motifs. As seen in 
Figure~\ref{fig:scenario_1_sampled}, the ISM tree requires 
site \( 147 \) to remain unmutated along the path \( \mathrm{E} \rightarrow \mathrm{F} \rightarrow \mathrm{S}_{1561} \) and the path \( \mathrm{E} \rightarrow \mathrm{F} \rightarrow \mathrm{S}_{4211} \), which is far less likely under the DSM. 
\paragraph{Scenario 2:}
In the example of Figure~\ref{fig:scenario_3_sampled}, the (sub)tree obtained under the ISM has Hamming distance 8 between the UCA (node $\mathrm{A}$) and $\mathrm{S}_{80}$. Since 4 of those 8 mutations are shared with $\mathrm{S}_{245}$, the ISM 
places $\mathrm{S}_{80}$ and $\mathrm{S}_{245}$ 
on the same lineage $A \rightarrow D$. However, the 
contexts of these 4 
mutations 
are highly mutable, and under the DSM
they are more likely to have been duplicated independently along 
separate lineages. 

\subsection{Analysis of HIV bnAb B cell repertoires}
\label{sec:HIVbnAbs}

We applied our method to infer the UCA and lineage tree of two well-characterized HIV bnAb clones, CH235 and DH270 which have served as prototypical examples of bnAb maturation as well as templates for lineage-based HIV vaccine design \cite{bonsignori_maturation_2016,Bonsignori_2017,Saunders_Wiehe_2019,Wiehe_2024}.  The DH270 clonal lineage tree was reconstructed using heavy chain sequences of 6 observed members and the CH235 clonal lineage was reconstructed using heavy chain sequences of 55 observed members. The limited number of observed sequences in the DH270 clone and the extensive diversity of CDRH3 sequences observed in the CH235 clone (see Figure~\ref{fig:CH235_observed} in Appendix) make these clones particularly challenging test cases for accurate clonal tree reconstruction.

\subsubsection{DH270}
\begin{figure}[htbp]
\centering
\begin{subfigure}[c]{.32\textwidth}
\centering
\resizebox{\textwidth}{!}{
\begin{tabular}{|c|c|}
    \hline
    Model & Estimated UCA \\
    \hline
    ISM & ATGGWI\textcolor{red}{G}LYYDSSGYPNFDY \\
    \hline
    DSM & ATGGWI\textcolor{red}{S}LYYDSSGYPNFDY \\
    \hline
\end{tabular}
}
\end{subfigure}
\hfill
\begin{subfigure}[c]{.65\textwidth}
\includegraphics[width=.9\textwidth,
trim = .1in .1in 0in .05in,clip]{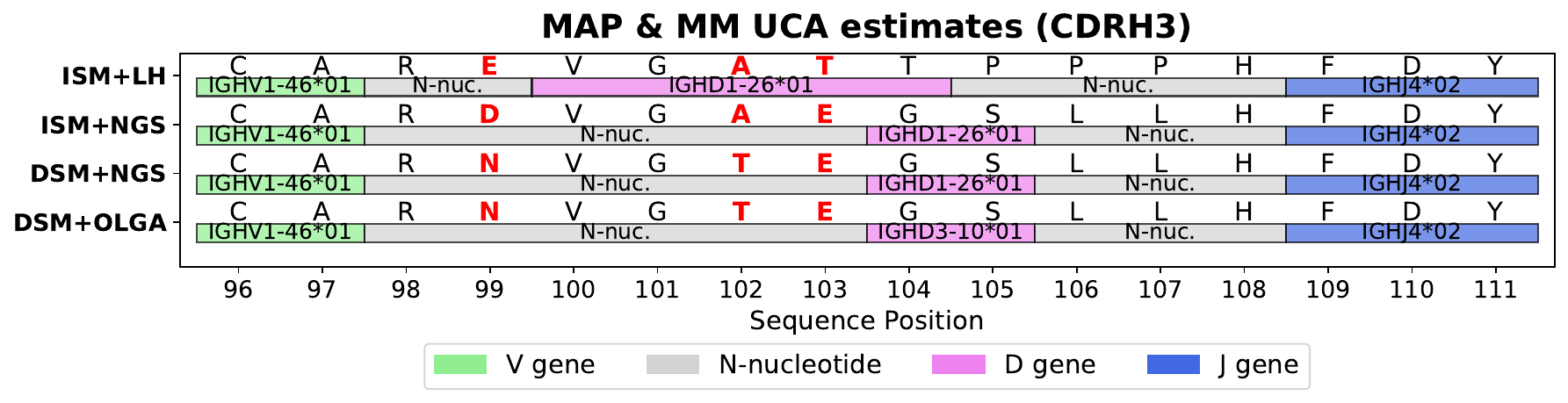}
\end{subfigure}
\caption{MAP UCA estimates for HIV bnAb DH270 and CH235 clones (displaying only CDR3 region). Shown are independent site model with VDJ prior using Linearham with (ISM+LH) default parameters,  vs. (ISM+NGS) parameters from a large human B cell NGS dataset; and context-dependent model with  (DSM+NGS) Linearham VDJ prior with NGS parameters), vs. (DSM+OLGA) OLGA~\cite{sethna_olga_2019} VDJ prior.}
\label{tab:MAP_DH270_UCA_CDRH3_Table}
\label{fig:CH235_UCA}
\end{figure}

The UCA estimated for the DH270 clone under the DSM differs by only a single amino acid from the UCA estimated under the ISM, in 7th position of CDRH3 
(Fig.~\ref{tab:MAP_DH270_UCA_CDRH3_Table}a, red). This site, which is reconstructed as a glycine under the ISM and a serine under the context-dependent model, occurs unambiguously in the non-templated nucleotide (n-nucleotide) regions that are not encoded by the V, D or J gene segments and thus cannot be resolved by priors on the initial VDJ rearrangement. While this difference in reconstructions would appear to be small, in fact the 7th position in the CRH3 is critical for binding of DH270 lineage members to the HIV Env protein as it makes structural contacts to the V3 glycan epitope \cite{Henderson2023}. Moreover, deep scanning mutagenesis in the CDRH3 of a previously inferred DH270 UCA revealed that the 7th position in the CDRH3 could only tolerate a serine to maintain binding to the designed priming immunogen ~\cite{swanson_identification_2023}. Thus, the one position difference between the ISM and DSM reconstructions occurs at a key site for both bnAb epitope recognition and for binding to an immunogen designed to activate the DH270 lineage, and represents a critical error of the ISM assumption. This example underscores the importance of highly accurate UCA estimates for both understanding the natural maturation history of clonal lineages and for vaccine design. While the true UCA sequence for the DH270 clone is unknowable, the fact that the DSM-estimated UCA contains the serine at this critical 7th position which is the only amino acid compatible with binding to the designed priming immunogen -- itself a modified form of an HIV Env circulating at the time the DH270 clone emerged -- suggests that the DSM-estimated UCA may more accurately reflect the true ancestral sequence. Moreover, designing an immunogen to elicit this bnAb based on the incorrect ISM-reconstructed UCA could be counterproductive, potentially selecting for mutants that compete \textit{against} the desired bnAb.

\begin{figure}[htbp]
\centering
\begin{subfigure}[b]{\linewidth}
\centering
\includegraphics[width=.95\linewidth,
trim = .4in .6in .4in .4in,clip]{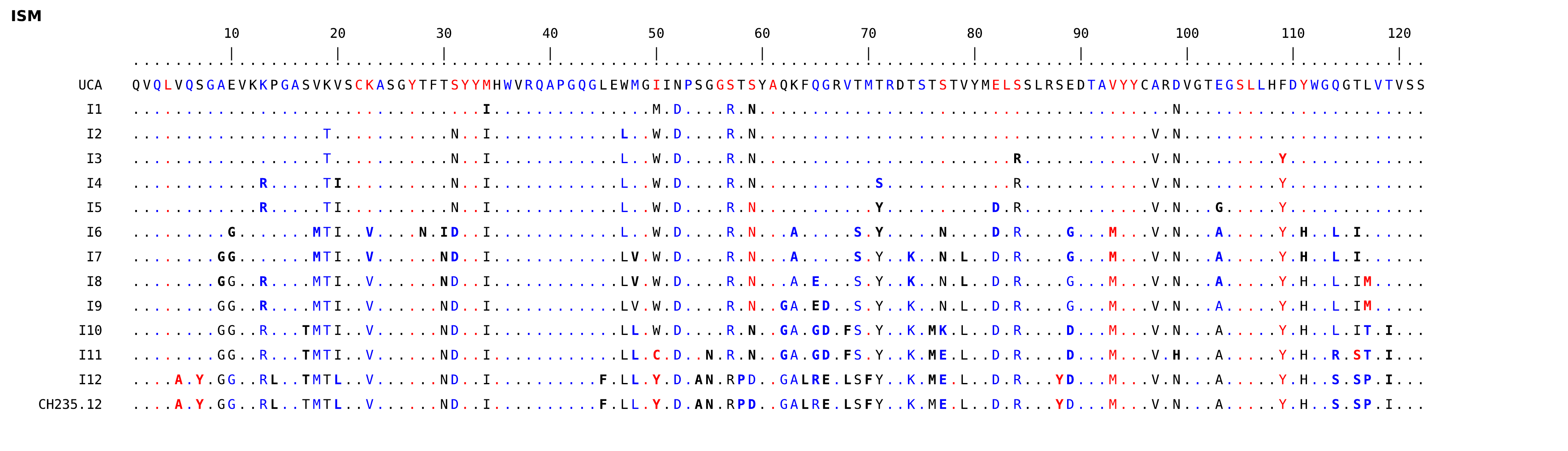}
\caption{ISM-reconstructed pathway to the HIV bnAb  CH235.12.
}
\label{fig:CH235_bnab_ISM}
\end{subfigure}\\
\begin{subfigure}[b]{\linewidth}
\centering
\includegraphics[width=.95\linewidth,
trim = .4in .6in .4in .4in,clip]{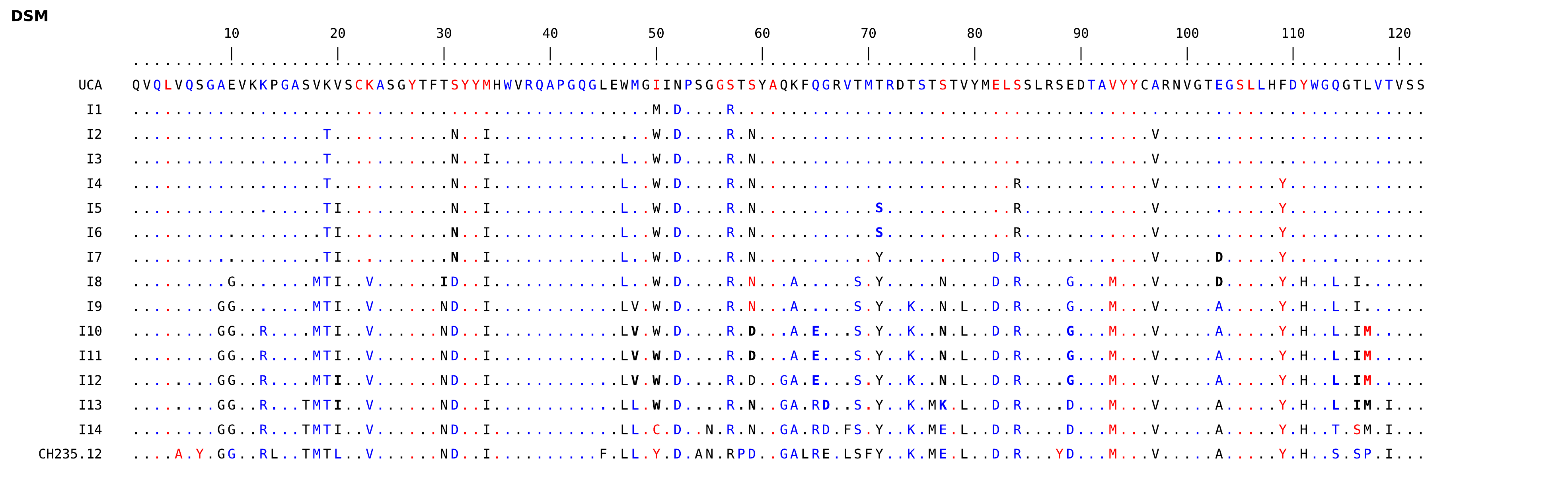}
\caption{DSM-reconstructed pathway to the HIV bnAb  CH235.12.
}
\label{fig:CH235_bnab_DSM}
\end{subfigure}
\caption{Reconstructed mutational pathways to an HIV bnAb. (Trees shown in Figure~A\ref{fig:CH235_DSM_NGS}.)
Red/blue 
indicates hot/cold spots in the codon, while bold characters highlight differing estimates between the models.}
\label{fig:CH235_bnab}
\end{figure}

\subsubsection{CH235} We observe two amino acid differences between the ISM and DSM reconstructions of the 
CH235 UCA CDRH3, at positions 99 and 102  (Fig.~\ref{fig:CH235_UCA}b). The CH235 CDRH3 forms extensive structural contacts with the epitope and is critical for recognition of the conserved CD4 binding site \cite{Henderson_2024}. Position 102 makes direct contact with the CH235 epitope, and position 99 likely influences CDRH3 positioning which may be critical for epitope recognition.  This underscores the importance of accurate inference of the unmutated CDRH3 in order to identify Env variants capable of engaging CH235 precursor B cells, a key step in immunogen design. 

Understanding the order in which critical mutations are acquired by bnAbs while developing neutralization breadth is also important for understanding how bnAb B cells mature during infection, and can inform vaccine strategies aiming to induce similar evolutionary trajectories in vaccinees \cite{Wiehe_2024}. Accurate estimates of internal nodes in the clonal lineage tree are therefore also important.
Observed differences in the reconstructed topologies under the DSM and ISM (Figure~\ref{fig:CH235_NGS}) indicate that the tree reconstruction is heavily influenced by the choice of model, with the  MAP trees showing key differences in mutational orderings (Figures~\ref{fig:CH235_bnab} \&
\ref{fig:CH235_NGS}), including differences in estimated acquisition timing 3 of mutations critical for bnAb epitope recognition. 
In the DSM tree, the W47L mutation which is critical for broad neutralization but highly improbable \cite{Wiehe:2018}, emarges later in the maturation pathway, 
suggesting it 
requires prolonged clonal residence in the germinal center to arise. Another mutation S59N, which contacts the Env V5 loop  in the CD4 binding site, is also estimated to occur later in the DSM vs. ISM MAP trees. These findings have implications for vaccine regimen design, as sequential boosting immunogens targeting key mutations may need to be timed carefully as boosting too early could misdirect maturation from its natural trajectory.
The ISM tree also displays three amino acid transitions at contact position 65 (Q$\to$E$\to$G$\to$R), whereas the DSM tree displays only two (Q$\to$E$\to$R), indicating that context-aware estimates can lead to more parsimonious lineages at critical positions. Together, these differences underscore the importance of modeling context-dependence in obtaining accurate inferences on the ordering and timing of mutations, both for understanding how B cells evolve and for designing vaccine strategies that aim to guide B cell maturation.

\section{Discussion}
The algorithms introduced here enable practical  reconstruction of B cell lineages and UCAs under context-dependent models of somatic hypermutation such as \Arm. 
Our results demonstrate that 
ignoring this context dependence, as all existing B cell lineage reconstruction methods do to our knowledge,
introduces errors that may have important impacts on lineage-based vaccine design.
We note that our  comparisons on simulated and real data show differences between ISM- and DSM-reconstructed UCAs at 
only a few sites.  This is because (a) existing VDJ models recover the (germline-encoded) majority of the BCR sequence with high accuracy, and (b) for moderate-to-large clones,  consensus reconstruction suffices for most $n$-nucleotide insertion sites.  However, reconstruction of tree topology and mutation ordering appears to be harder --  and more senstive to context dependence --  than UCA reconstruction, with the DSM performing significantly better (Fig.~\ref{fig:ism_pls},\ref{fig:dsm_pls}), and UCA differences are often attributable to the ISM getting the tree partially wrong. For the HIV bnAb clones analyzed the UCA differences, while few,  are functionally critical, while the observed differences in mutation ordering have real consequences for vaccine design.

Both the SS and SMC algorithms described here are highly parallelizable. The SMC approach is currently preferred due to the high variance of importance weights observed for SS;
as described in Appdx~\ref{sec:AppdxVar} this is due to the sampled $(\x,Z)$ distribution.  We are currently exploring whether sampling from improved approximations to $\pi_1(\x,Z \mid Y,g)$ based on pruning with truncated dependence can make this method competitive with SMC.  Another significant limitation of our importance-sampling approach is that we currently only evaluate 
$\pi_1(g \mid Y)$ for trees drawn from $\pi_0(g \mid Y)$.  Supplementing sampled trees by performing a local search 
around large weight trees (e.g. \cite{whidden_systematic_2020}) may further improve topology exploration.

\clearpage
\newpage
\bibliographystyle{splncs04}
\bibliography{reference}

\appendix
\setcounter{figure}{0}
\renewcommand{\thefigure}{S\arabic{figure}}

\section{Pairwise MCMC kernel}
\label{Sec:AppdxA}
Denote the unobserved \textit{full sequence} path as $\bPa = ( \Pa_1, \ldots, \Pa_n )$, where the path at the $i$-th site
\begin{align*}
\Pa_i = ( m_i,t_{\text{start}}, t^1_i,\ldots,t^{m_i}_i, t_{\text{end}}, b^0_i, b^1_i,\ldots,b^{m_i}_i )
\end{align*}
is specified by the number of mutations $m(\Pa_i) = m_i \in \{0,1,\ldots \}$  occurring at site $i$ along the path, the times $t^1_i, \ldots, t^{m_i}_i \in \mathbb{R}_{+}$ of these mutation events (satisfying $ t_{\text{start}} = t^0_i < t^1_i < \ldots < t^{m_i}_i < t^{m_{i}+1}_i = t_{\text{end}}$), and the identities $b^1_i,\ldots,b^{m_i}_i \in \mathscr{A} $ of the base changes. Note that both \( \Pa_{i} : [t_{\text{start}}, t_{\text{end}} ] \rightarrow \mathscr{A} \) and \( {\Tilde{x}}_{i} : [t_{\text{start}}, t_{\text{end}} ] \rightarrow {\mathscr{A}}^{3} \) can be viewed as functions, representing the character and the context, respectively, at a specific time stamp. We desire to sample \( P_{\Tilde{\mathbf{Q}}} ( \Pa_{i} \mid \bPa_{[-i]}, \x, \y ) \).

For the unobserved full sequence path $\bPa$, we can write the the density under the ISM defined by rates $\bQ$ as
\begin{equation*}
    P_{\bQ} ( \y, \bPa \mid \x ) = \prod_{i=1}^{\ell} {\bar{P}}_{\bQ} ( \Pa_{i} )
\end{equation*}
where
\begin{equation*}
    {\bar{P}}_{\bQ} ( \Pa_{i} ) = \prod_{j=1}^{m_{i}} \gamma (b_{i}^{j}; b_{i}^{j-1}) \prod_{j=0}^{m_{i}} e^{- (t_{i}^{j+1} - t_{i}^{j} ) \sum_{b \in \mathscr{A}}\gamma(b; b_{i}^{j})}.
\end{equation*}
Similarly, the density for \( \bPa \) under the DSM defined by rate \( \Tilde{\bQ} \) is given by
\begin{equation*}
    P_{\Tilde{\bQ}} ( \y, \bPa \mid \x ) = \prod_{j=1}^{\ell} {\bar{P}}_{\Tilde{\bQ}} ( \Pa_{i} \mid {\Tilde{x}}_{i}, {\Tilde{y}}_{i}, \Pa_{i-1}, \Pa_{i+1} )
\end{equation*}
where
\begin{equation*}
     {\bar{P}}_{\Tilde{\bQ}} ( \Pa_{i} \mid {\Tilde{x}}_{i}, {\Tilde{y}}_{i}, \Pa_{i-1}, \Pa_{i+1} ) = \prod_{j=1}^{m_{i}} \Tilde{\gamma} (b_{i}^{j}; {\Tilde{x}}_{i} ( t_{i}^{j} ) ) \prod_{j=0}^{m_{i}} \exp \left\{- \int_{t_{\text{start}}}^{t_{\text{end}}} \Tilde{\gamma} ( \cdot; {\Tilde{x}}_{i} ( t ) ) dt \right\}.
\end{equation*}

Let $\bar{m}^{\prime}_i := m_{i+1} + m_{i-1}$ be the total number of jumps among the paths $\Pa_{i-1}$ and $\Pa_{i+1}$, and $ t_{\text{start}} = t^0 < t^1 < t^2 < \ldots < t^{\bar{m}^{\prime}_i} < t_{\text{end}}$  the corresponding jump times. Denote \( {\bar{t}}_{i} = \left\{ t^{1}, \ldots, t^{\bar{m}^{\prime}_i} \right\} \). The Metropolis-Hastings chain proposes an updated path $\Pa_i^{\prime}$ at site $i$ conditional on the neighboring site paths $\Pa_{i-1}$ and $\Pa_{i+1}$ using the following algorithm:
\paragraph{\underline{\textit{Hobolth algorithm}}:}
\begin{enumerate}
\item Simulate the bases ${\bar{\mathbf{x}}}_{i} = ( x_i(t^1), \ldots,x_i(t^{\bar{m}^{\prime}_i}) )$ at site $i$ at the neighboring jump times using forward-filtering/backward-sampling.
\item Form the sample path $\Pa_i = (\Pa_{i,0} : \ldots : \Pa_{i,\bar{m}^{\prime}_i})$, where $\Pa_{i,j}$ is a `subpath' with endpoints $x_i(t^{j})$ and $x_i(t^{j+1})$ sampled using single-site endpoint-conditioned sampling techniques \cite{hobolth_markov_2008}.
\end{enumerate}
Hobolth algorithm is effectively proposing paths from a position-dependent time-inhomogeneous ISM. In particular, for position \( i \) fixing the neighboring path \( \Pa_{i-1} \) and \( \Pa_{i+1} \), the ISM is defined by rates
\begin{equation*}
    \bQ_{ab} ( t \mid \Pa_{i-1}, \Pa_{i+1} ) = {\Tilde{\gamma}} (b; ( \Pa_{i-1} ( t ), a, \Pa_{i+1} ( t ) ) ).
\end{equation*}
Note that this CTMC is time-homogeneous within time the interval \( ( t^{j}, t^{j+1} ) \) for each \( i \in \{ 0, 1, \ldots, {\bar{m}}_{i}' \} \). Let \( \bQ^{(i,j)} \) denote the rate matrix for position \( i \) within the interval \( ( t^{j}, t^{j+1} ) \). Therefore, the density of the characters drawn in step 2, \( q ( {\bar{\mathbf{x}}}_{i} \mid {\Tilde{x}}_{i}, {\Tilde{y}}_{i}, {\bar{t}}_{i}, \Pa_{i-1}, \Pa_{i+1} ) \), can be calculated by belief propagation for time-inhomogeneous CTMC.
The resulting density of the path drawn in step 3 is given by
\begin{equation*}
    q (\Pa_{i} \mid {\Tilde{x}}_{i}, {\Tilde{y}}_{i}, {\bar{t}}_{i}, \Pa_{i-1}, \Pa_{i+1}) = \sum_{{\bar{\mathbf{x}}}_{i} \in {\mathscr{A}}^{{\bar{m}}_{i}'}} q ( {\bar{\mathbf{x}}}_{i} \mid {\Tilde{x}}_{i}, {\Tilde{y}}_{i}, {\bar{t}}_{i}, \Pa_{i-1}, \Pa_{i+1} ) \prod_{j=0}^{m_{i}} P_{\bQ^{(i,j)}} ( x_{i} ( t^{j+1} ), \Pa_{i} (t^{j} : t^{j+1} ) \mid x_{i} ( t^{j} ) ),
\end{equation*}
which can be viewed as a mixture model wherein sampling \( {\bar{\mathbf{x}}}_{i} \) selects the mixture component. Plugging in \( P_{\bQ^{(i, j)}} \) gives:
\begin{equation*}
    q (\Pa_{i} \mid {\Tilde{x}}_{i}, {\Tilde{y}}_{i}, {\bar{t}}_{i}, \Pa_{i-1}, \Pa_{i+1}) = P_{\Tilde{Q}} ( \Pa_{i} \mid {\Tilde{x}}_{i}, {\Tilde{y}}_{i}, \Pa_{i-1}, \Pa_{i+1}),
\end{equation*}
where the corresponding density of sampling \( {\bar{\mathbf{x}}}_{i} \) does not appear.
Recognizing that we desire to sample
\begin{equation*}
    p_{1} ( \Pa_{i} \mid \x, \y, \Pa_{[-i]}) \propto P_{\Tilde{\bQ}} ( \y, \bPa \mid \x ) = \prod_{j=1}^{\ell} {\bar{P}}_{\Tilde{\bQ}} ( \Pa_{i} \mid {\Tilde{x}}_{i}, {\Tilde{y}}_{i}, \Pa_{i-1}, \Pa_{i+1} )
\end{equation*}
where \( \Pa_{[-i]} := ( \Pa_{1}, \ldots, \Pa_{i-1}, \Pa_{i+1}, \ldots, \Pa_{\ell} )\), the acceptance ratio of the Metropolis-Hastings chain is given by
\begin{equation*}
a ( \bPa', \bPa ) := \min\left\{1, \frac{{\bar{P}}_{\Tilde{\bQ}} ( \Pa_{i-1} \mid {\Tilde{x}}_{i-1}, {\Tilde{y}}_{i-1}, \Pa_i', \Pa_{i-2}) {\bar{P}}_{\Tilde{\bQ}} ( \Pa_{i+1} \mid {\Tilde{x}}_{i+1}, {\Tilde{y}}_{i+1}, \Pa_i', \Pa_{i+2} )}{{\bar{P}}_{\Tilde{\bQ}} ( \Pa_{i-1} \mid {\Tilde{x}}_{i-1}, {\Tilde{y}}_{i-1}, \Pa_i, \Pa_{i-2}) {\bar{P}}_{\Tilde{\bQ}} ( \Pa_{i+1} \mid {\Tilde{x}}_{i+1}, {\Tilde{y}}_{i+1}, \Pa_i, \Pa_{i+2} )} \right\}.
\end{equation*}

\section{Forward-filtering/backward-sampling}
\label{Sec:FFBS}
Forward–filtering / backward–sampling (FFBS) efficiently draws latent intermediate states or internal states from a time-inhomogeneous CTMC with piecewise constant rates, as used in the Hobolth proposal.

\subsection{Sampling the Intermediate States} 
Recall that we desire to sample \( q ( \bar{\x}_{i} \mid \Tilde{x}_{i}, \Tilde{y}_{i}, \bar{t}_{i} \Pa_{i-1}, \Pa_{i+1} ) \) in the first step of Hobolth proposal. Recall that $\bar{m}^{\prime}_i $ and \( {\bar{t}}_{i} \) are the total number of jumps and the corresponding jump times among the paths $\Pa_{i-1}$ and $\Pa_{i+1}$, respectively. Denoting \( t^{0} = t_{\text{start}} \) and \( t^{\bar{m}_{i}'+1} = t_{\text{end}} \), we have
\begin{align*}
    q ( \bar{\x}_{i} \mid \Tilde{x}_{i}, \Tilde{y}_{i}, \bar{t}_{i} \Pa_{i-1}, \Pa_{i+1} ) & \propto q \left( x_{i} ( t^{1} ), \ldots, x_{i} ( t^{\bar{m}_{i}'} ), x_{i} ( t^{\bar{m}_{i}'+1} ) \mid x_{i} ( t^{0} ), x_{i-1}, x_{i+1}, y_{i-1}, y_{i+1}, \Pa_{i-1}, \Pa_{i+1} \right) \\
    & = \prod_{j=0}^{\bar{m}_{i}'} q \left( x_{i} ( t^{j+1} ) \mid x_{i} ( t^{0} ), \ldots, x_{i} ( t^{j} ), x_{i-1}, x_{i+1}, y_{i-1}, y_{i+1}, \Pa_{i-1}, \Pa_{i+1} \right) \\
    & \stackrel{*}{=} \prod_{j=0}^{\bar{m}_{i}'} q \left( x_{i} ( t^{j+1} ) \mid x_{i} ( t^{j} ), x_{i-1}, x_{i+1}, y_{i-1}, y_{i+1}, \Pa_{i-1}, \Pa_{i+1} \right) \\
    & = \prod_{j=0}^{\bar{m}_{i}'} {\left( e^{\bQ^{(i, j)}} \right)}_{x_{i} ( t^{j} ) x_{i} ( t^{j+1} )} 
\end{align*}
where \( (*) \) is invoking the Markov property. We therefore construct Algorithm~\ref{alg:ffbs} to sample the intermediate states at a site given its neighboring paths according to the Hobolth proposal.

\begin{algorithm2e}[H]
\caption{FFBS for sampling intermediate states of site \(i\)}
\label{alg:ffbs}
\DontPrintSemicolon
\LinesNumbered

\KwIn{Endpoints \(\Tilde{x}_i \), \(\Tilde{y}_i\); neighboring paths \(\Pa_{i-1},\Pa_{i+1}\); rate matrices \(\{\bQ^{(i,j)}\}\).}
\KwOut{Sequence of intermediate states 
\(\bar\x_i=\{x_i(t^1),\ldots,x_i(t^{\bar m_i'})\}\).}

Initialize transition matrices

\For(\tcp*[h]{Initialize transition matrices}){$ j \leftarrow 0 $ \KwTo $ {\bar{m}}_{i}' $}{
Calculate \( \bP^{(i,j)} = e^{\bQ^{(i,j)}} \).}

\For(\tcp*[h]{Forward filtering}){$ j \leftarrow 1 $ \KwTo $ {\bar{m}}_{i}' $}{
Calculate \(
\upsilon_{j}(b)=\sum_{a\in\mathscr{A}}\upsilon_{j-1}(a)\bP^{(i,j-1)}_{ab}.
\)
}

\For(\tcp*[h]{Backward sampling}){$j=\bar{m}'_{i}$ \KwTo $1$}{
Calculate \(
\Pr(x_i(t^{j})=a\mid x_{i} ( t^{0} ), x_i(t^{j+1})=b)
=\frac{\upsilon_{j}(a)\bP^{(i,j)}_{ab}}
{\sum_{a'}\upsilon_{j}(a')\bP^{(i,j)}_{a'b}}.
\)\;
Sample \(x_i(t^{j})\) accordingly.
}

\end{algorithm2e}

\subsection{Sampling the Internal State}

We desire to sample \( q ( s_{i} \mid \Tilde{x}_{i}, \Tilde{y}_{i}, \Tilde{z}_{i}, \bPa^{(\x\s\y\z)}_{i-1}, \bPa^{(\x\s\y\z)}_{i+1} ) \).
\begin{align}
    q \left( s_{i} \mid \Tilde{x}_{i}, \Tilde{y}_{i}, \Tilde{z}_{i}, \bPa^{(\x\s\y\z)}_{i-1}, \bPa^{(\x\s\y\z)}_{i+1} \right) & \propto q \left( s_{i}, y_{i}, z_{i} \mid x_{i}, x_{i-1}, x_{i+1}, y_{i-1}, y_{i+1}, z_{i-1}, z_{i+1}, \bPa^{(\x\s\y\z)}_{i-1}, \bPa^{(\x\s\y\z)}_{i+1} \right) \nonumber \\
    & = q \left( s_{i} \mid \Tilde{x}_{i}, \bPa^{(\x\s)}_{i-1}, \bPa^{(\x\s)}_{i+1} \right) q \left( y_{i}, z_{i} \mid s_{i}, \bPa^{(\s\y)}_{i-1}, \bPa^{(\s\y)}_{i+1}, \bPa^{(\s\z)}_{i-1}, \bPa^{(\s\z)}_{i+1} \right) \nonumber \\
    & = q \left( s_{i} \mid \Tilde{x}_{i}, \bPa^{(\x\s)}_{i-1}, \bPa^{(\x\s)}_{i+1} \right) q \left( y_{i} \mid \Tilde{s}_{i}, \bPa^{(\s\y)}_{i-1}, \bPa^{(\s\y)}_{i+1} \right) q \left( z_{i} \mid \Tilde{s}_{i}, \bPa^{(\s\z)}_{i-1}, \bPa^{(\s\z)}_{i+1} \right), \label{eq:start-tree}
\end{align}
where the second equality comes from the conditional independence between \( \y \) and \( \z \) given \( \s \). Recall that Hobolth algorithm is effectively a time-inhomogeneous CTMC. Using the notations \( \bar{m}_{i}' \) and \( \bQ^{(i, j)} \) for the pairwise paths \( \bPa^{(\x\s)}_{i-1} \) and \( \bPa^{(\x\s)}_{i+1} \), we have
\begin{equation*}
    q \left( s_{i} \mid \Tilde{x}_{i}, \bPa^{(\x\s)}_{i-1}, \bPa^{(\x\s)}_{i+1} \right) = {\left( \prod_{j=0}^{\bar{m}_{i}'} e^{\bQ^{(i, j)}} \right)}_{x_{i} s_{i}}.
\end{equation*}
Similar calculation can be performed for the latter two terms in \eqref{eq:start-tree}, giving rise to Algorithm~\ref{alg:ffbsStar} for sampling the internal state of a star-tree.

\begin{algorithm2e}[H]
\caption{Sampling the internal state \(s_i\)}
\label{alg:ffbsStar}
\DontPrintSemicolon
\LinesNumbered

\KwIn{Endpoints \(\Tilde{x}_i,\Tilde{y}_i,\Tilde{z}_i\); neighboring paths 
\(\bPa^{(\x\s\y\z)}_{i-1}, \bPa^{(\x\s\y\z)}_{i+1}\);
rate matrices \(\{\bQ^{(i,j)}\}\) for pairwise paths.}
\KwOut{Internal state \(s_i\).}

Compute pairwise transition probability \( q \left( s_{i} \mid \Tilde{x}_{i}, \bPa^{(\x\s)}_{i-1}, \bPa^{(\x\s)}_{i+1} \right) \).

Compute pairwise transition probability \( q \left( s_{i} \mid \Tilde{y}_{i}, \bPa^{(\s\y)}_{i-1}, \bPa^{(\s\y)}_{i+1} \right) \).

Compute pairwise transition probability \( q \left( s_{i} \mid \Tilde{z}_{i}, \bPa^{(\s\z)}_{i-1}, \bPa^{(\s\z)}_{i+1} \right) \).

\ForEach(\tcp*[h]{Calculate unnormalized conditional probability}){\( s_i \in \mathscr{A} \)}{
\(
f(s_i) =
q \left( s_{i} \mid \Tilde{x}_{i}, \bPa^{(\x\s)}_{i-1}, \bPa^{(\x\s)}_{i+1} \right) q \left( s_{i} \mid \Tilde{y}_{i}, \bPa^{(\s\y)}_{i-1}, \bPa^{(\s\y)}_{i+1} \right) q \left( s_{i} \mid \Tilde{z}_{i}, \bPa^{(\s\z)}_{i-1}, \bPa^{(\s\z)}_{i+1} \right) \).
}
\ForEach(\tcp*[h]{Normalization}){\( s_i \in \mathscr{A} \)}{\( \mathrm{Pr}(s_i) = f(s_i)/\sum_{s'_i}f(s'_i) \).}

Sample \(s_i\) from the categorical distribution \(\mathrm{Pr}(s_i)\).\;
\end{algorithm2e}
\section{Star-path MCMC kernel} \label{Sec:AppdxStartPath}
We introduce a Markov kernel on tree-path space which updates tree-paths component-wise by (randomly or systematically) selecting a tree node $\s$ 
and proposing  a \textit{star-path} change; see Figure~\ref{fig:example}. Each of the four nodes represents a sequence of length $\ell$, e.g. $ \mathbf{x} = \{ x_1, x_2, \ldots, x_{\ell - 1}, x_{\ell} \}$. Let $\s_{-i} = \{ s_1, \ldots, s_{i-1}, s_{i+1}, \ldots, s_{\ell} \}$, and let
\begin{figure}[htbp]
\centering
\begin{tikzpicture}[node distance = 15mm and 15mm,
V/.style = {circle, draw, fill=gray!30}, every edge quotes/.style = {auto, font=\footnotesize, sloped}]
\begin{scope}[nodes=V]
    \node (x) at (0, 0) {\( \mathbf{x} \)};
    \node (s) at (0, -1.2) {\( \mathbf{s} \)};
    \node (y) at (-1, -2.2) {\( \mathbf{y} \)};
    \node (z) at (1, -2.2) {\( \mathbf{z} \)};
\end{scope}

\draw[->, thick] (x) edge[""] (s)
    (s) edge[""] (y)
    (s) edge[""] (z);
\end{tikzpicture}
\caption{Example}
\label{fig:example}
\end{figure}
\noindent $\bPa^{(\x\s)} := (\Pa^{(\x\s)}_{1}, \ldots, \Pa^{(\x\s)}_{\ell} )$ denote an evolutionary path from $\x$ to $\s$, where $\Pa^{(\x\s)}_j$ is the evolutionary path at site $j$ (from $x_j$ to $s_j$). We update the \textit{single-site star-path} $ \bPa^{(\x\s\y\z)}_{i} := ( \Pa^{(\x\s)}_{i}, \Pa^{(\s\y)}_{i}, \Pa^{(\s\z)}_{i} )$ using a Metropolis-Hastings step as follows:
\begin{enumerate}
\item Sample character $s_i'$ from the modified Hobolth proposal $q ( \cdot )$ (see Appendix) conditioning on 
$\bPa^{(\x\s\y\z)}_{i-1}$ and $\bPa^{(\x\s\y\z)}_{i+1} $:
\begin{equation*}
s_i' \sim q ( \cdot \mid {\Tilde{x}}_i, {\Tilde{y}}_i, {\Tilde{z}}_i, \bPa^{(\x\s\y\z)}_{i-1}, \bPa^{(\x\s\y\z)}_{i+1} ).
\end{equation*}
and set $\s' = (s_1,\ldots,s_{i-1},s_i',s_{i+1}, \ldots s_{\ell}$).
\item Draw the star-path $\bPa^{(\x\s'\y\z)}_i $ from the Hobolth proposal for pairwise paths (see Appendix~\ref{Sec:AppdxA}), i.e.
\begin{align*}
\Pa^{(\x\s')}_{i} & \sim q ( \cdot \mid \Tilde{x}_i, \Tilde{s}_i', \bar{t}^{(\x\s)}_{i}, \Pa^{(\x\s)}_{i-1}, \Pa^{(\x\s)}_{i+1} )\\
\Pa^{(\s'y)}_i & 
\sim q ( \cdot \mid \Tilde{s}_i', \Tilde{y}_i, \bar{t}^{(\s\y)}_{i}, \Pa^{(\s\y)}_{i-1}, \Pa^{(\s\y)}_{i+1} )\\
\Pa^{(\s'\z)}_i & \sim q ( \cdot \mid \Tilde{s}_i', \Tilde{z}_i, \bar{t}^{(\s\z)}_{i}, \Pa^{(\s\z)}_{i-1}, \Pa^{(\s\z)}_{i+1} )
\end{align*}
%
where \( \bar{t}^{(\x\s)}_{i} \) are the times of the jumps among the paths $\Pa^{(\x\s)}_{i-1}$ and $\Pa^{(\x\s)}_{i+1}$,
%
since $\Pa^{(\x\s')}_i, \Pa^{(\s'\y)}_i$ and $\Pa^{(\s'\z)}_{i}$ are conditionally independent of each other given $\Tilde{x}_i, \Tilde{y}_i, \Tilde{z}_i, \bPa^{(\x\s\y\z)}_{i-1}, \bPa^{(\x\s\y\z)}_{i+1}$ and $s_i'$ under \( q ( \cdot ) \).
\item Accept \( \bPa^{(\x\s'\y\z)}_i \) with probability
\begin{equation*}
\alpha ( \bPa^{(\x\s'\y\z)}, \bPa^{(\x\s\y\z)} ) := \min \left\{ 1, a ( \bPa^{(\x \s')},  \bPa^{(\x\s)} ) a ( \bPa^{(\s'\y)},  \bPa^{(\s\y)} ) a ( \bPa^{(\s'\z)},  \bPa^{(\s\z)} ) \right\},
\end{equation*}
where $a ( \cdot, \cdot )$ is the acceptance ratio of the pairwise MCMC kernel given in Appendix~\ref{Sec:AppdxA}.
\end{enumerate}
Note that $q$ can be viewed as a finite mixture distribution wherein sampling $s_i'$ selects the mixture component:
\begin{align*}
    & q ( \bPa^{(\x\s'\y\z)}_{i} \mid {\Tilde{x}}_{i}, {\Tilde{y}}_{i}, {\Tilde{z}}_{i}, \bPa^{(\x\s'\y\z)}_{i-1}, \bPa^{(\x\s'\y\z)}_{i+1} ) \\
    = & \sum_{s_{i}' \in \mathscr{A}} q ( s_{i}' \mid {\Tilde{x}}_i, {\Tilde{y}}_i, {\Tilde{z}}_i, \bPa^{(\x\s\y\z)}_{i-1}, \bPa^{(\x\s\y\z)}_{i+1} ) q ( \bPa^{(\x\s'\y\z)}_{i} \mid {\Tilde{x}}_{i}, \Tilde{s}_{i}', {\Tilde{y}}_{i}, {\Tilde{z}}_{i}, \bPa^{(\x\s'\y\z)}_{i-1}, \bPa^{(\x\s'\y\z)}_{i+1} )
\end{align*}
where \( q ( \bPa^{(\x\s'\y\z)}_{i} \mid {\Tilde{x}}_{i}, \Tilde{s}_{i}', {\Tilde{y}}_{i}, {\Tilde{z}}_{i}, \bPa^{(\x\s'\y\z)}_{i-1}, \bPa^{(\x\s'\y\z)}_{i+1} ) \) factors as
\begin{equation*}
    q ( \Pa^{(\x \s')}_{i} \mid \Tilde{x}_i, \Tilde{s}_i', \bar{t}^{(\x\s)}_{i}, \Pa^{(\x\s)}_{i-1}, \Pa^{(\x\s)}_{i+1} ) q ( \Pa^{(\s'\y)}_{i} \mid \Tilde{s}_i', \Tilde{y}_i, \bar{t}^{(\s\y)}_{i}, \Pa^{(\s\y)}_{i-1}, \Pa^{(\s\y)}_{i+1} ) q ( \Pa^{(\s'\z)}_{i} \mid \Tilde{s}_i', \Tilde{z}_i, \bar{t}^{(\s\z)}_{i}, \Pa^{(\s\z)}_{i-1}, \Pa^{(\s\z)}_{i+1} ).
\end{equation*}
Hence the density of sampling $s_i'$ does not appear in the Metropolis-Hastings acceptance probability.

\section{UCA Estimators} \label{Sec:AppedxUCAEst} 
As described above, the marginal joint posterior distribution of UCA $\x$ and genealogy $g$ under the ISM is given by
\begin{equation*}
p_0(\x, g \mid Y) 
= p_0(\x \mid Y, g) \pi_0(g \mid Y)
\end{equation*}
which can be sampled by using standard Bayesian phylogenetics packages to draw $g \sim \pi_0(\cdot \mid Y)$ and then applying Felsenstein's pruning to draw $\x \sim p_0(\cdot \mid Y, g)$.
The corresponding importance weight of interest
\begin{equation}
\frac{p_1(\x,g \mid Y)}{p_0(\x,g \mid Y)} \propto \frac{p_1 (Y \mid \x, g) \zeta_1 (\x)}{p_0 (Y \mid \x, g) \zeta_0 (\x)} =: w(\x, g) 
\label{eq:UCAweight1}
\end{equation}
can then be approximated by the following unbiased Monte Carlo estimator:
\begin{equation} 
\hat{w}(\x, g) = \frac{1}{m} \sum_{j=1}^m \frac{p_1(Y, Z^{(j)} \mid \x, g)}{p_0(Y, Z^{(j)} \mid \x, g)}
\label{eq:UCAweight}
\end{equation}
Given samples \(\big( \x^{(i)}, Z^{(i)}, g^{(i)} \big)\) for $i=1,\ldots,q$ drawn from the joint posterior $p_0(\x,Z,g \mid Y)$, under the ISM and corresponding importance weights $\hat{w}(\x^{(i)}, g^{(i)})$, \eqref{eq:UCAweight} weight becomes
\begin{equation*}
\hat{w}_j \defeq \frac{\hat{w} \big(\x^{(j)}, g^{(j)} \big)}{\sum_{i=1}^{q} \hat{w} \big(\x^{(i)}, g^{(i)} \big)} \quad \text{where} \quad\hat{w} \big(\x^{(j)}, g^{(j)} \big) = \sum_{\{i: x^{(i)} = x^{(j)}\}} \hat{w}(\x^{(i)},Z^{(i)}, g^{(i)} ).
\end{equation*}
We consider two estimators of the UCA which can be obtained directly from $\hat{w}_j$:
\begin{align*}
\xmap 
\defeq \max_{x} \sum_{i=1}^{q} {\mathbbm{1}}_{\x = \x^{\left( i \right)}} {\hat{w}}_{i}
\qquad\quad 
\xmm
\defeq \left\{ \max_{x_{j} \in \left\{\mathrm{A}, \mathrm{C}, \mathrm{G}, \mathrm{T} \right\}} \sum_{i=1}^{q} {\mathbbm{1}}_{x_{j} = x_{j}^{\left( i \right)}} {\hat{w}}_{i} \right\},
\end{align*}
for \( {\mathbbm{1}}_{\{\}} \) an indicator function. 
$\xmap$ estimates the \textit{maximum a posteriori} UCA by the sampled UCA with the highest estimated posterior probability, while $\xmm$ estimates the marginal mode nucleotide at each sequence position independently by selecting the most probable nucleotide at each position. For comparison, supposing that \( \x^{*} \) is the true value of the UCA, we also report estimated expected posterior loss under the Hamming distance 
\begin{align*}
\hat{E}_{p_1}[d_H(\x,\x^\star)] 
=
    \frac{1}{q} \sum_{i=1}^{q} {\hat{w}}_{i} d_{H} \big( \x^{\left( i \right)}, \x^{*} \big)
    \qquad
\hat{E}_{p_0}[d_H(\x,\x^\star)] 
=    
     \frac{1}{q} \sum_{i=1}^{q} d_{H} \big( \x^{\left( i \right)}, \x^{*} \big). 
\end{align*}
where $d_{H}(\x,\y) = \sum^{n}_{i=1} \delta(x_{i} \neq y_{i})$ is the Hamming distance between sequences $\x$ and $\y$. The expected loss captures the accuracy of using the entire posterior distribution rather than a single point estimator.

\section{Variance Decomposition} \label{sec:AppdxVar}

By law of total variance, we can decompose the variance of the importance weights of tree-paths as
\begin{align}
    \text{Var} ( \frac{p_{1} ( Y, \boldsymbol{\mathcal{P}}^{(g)}, \x \mid g )}{p_{0} ( Y, \boldsymbol{\mathcal{P}}^{(g)}, \x \mid g )} ) = & {\mathbb{E}}_{p_{0} ( Z, \x \mid Y, g)} \left( \text{Var}_{p_{0} ( \bPa^{(g)} \mid Y, Z, \x, g )} \left( \frac{p_{1} ( Y, Z, \boldsymbol{\mathcal{P}}^{(g)}, \x \mid g )}{p_{0} ( Y, Z, \boldsymbol{\mathcal{P}}^{(g)}, \x \mid g )} \mid Z, \x \right) \right) \nonumber \\
    + & \text{Var}_{p_{0} ( Z, \x \mid Y, g )} \left( {\mathbb{E}}_{p_{0} ( \bPa^{(g)} \mid Y, Z, \x, g )} \left( \frac{p_{1} ( Y, Z, \boldsymbol{\mathcal{P}}^{(g)}, \x \mid g )}{p_{0} ( Y, Z, \boldsymbol{\mathcal{P}}^{(g)}, \x \mid g )} \mid Z, \x \right) \right) \nonumber \\
    = & \mathbb{E} \left( \text{Var} \left( \frac{p_{1} ( Y, Z, \boldsymbol{\mathcal{P}}^{(g)}, \x \mid g )}{p_{0} ( Y, Z, \boldsymbol{\mathcal{P}}^{(g)}, \x \mid g )} \mid Z, \x \right) \right) + \text{Var}_{p_{0} ( Z, \x \mid Y, g )} \left( \frac{p_{1} ( Y, Z, \x \mid g )}{p_{0} ( Y, Z, \x \mid g )} \right). \label{eq:VarDecom}
\end{align}
The first component is the expected conditional variance due to the stochasticity in the path samples given the internal sequence \( \x \) and \( Z \). The second component is the variance of the conditional expectation, the variance across the different possible internal sequence configurations. One option to control the variance is to condition more tightly on the internal sequences, which motivates the stratified sampling.

Our stratified sampling algorithm reduces variance by uniformly (over $\x$ and $Z$) bounding the first variance 
term in~\eqref{eq:MargLikEdgeProduct}, which is conceptually similar to Rao-Blackwellization. However, the stratified sampling does not control the second term in~\eqref{eq:VarDecom}, which relates to the distance between the marginal joint distributions of $\x,Z \mid g, Y$ under the ISM vs DSM distributions.

At first glance, Table~\ref{Tbl:ESSComparison} appears to suggest that the stratified sampling strategy yields higher variance than standard importance sampling, which would seem to contradict 
the above analysis. 
However, this is not the case, as the reported ESS for stratified sampling corresponds to roughly 1.11 effective samples of \textit{internal sequences $(\x,Z)$}. Since the first term in
\eqref{eq:VarDecom} can be made arbitrarily small using the edge-weight estimator in~\cite{Mathews:2024}, 
the remaining variance primarily arises from the distance between the instrumental (ISM) and target (DSM) distributions. This observation motivates the sequential Monte Carlo strategy that progressively bridges these two distributions.

\section{Experiments}
\label{sec:AppExp}

\paragraph{\textbf{Model Parameters}} For the DSM, we consistently use the mutation rates specified by the \Arm model \cite{Wiehe:2018}. For the ISM, we employ the corresponding mean-field approximation. This choice helps reduce the possibility that differences in the inferred trees or UCAs between the ISM and DSM arise from disparities in the underlying rate parameters rather than from the presence or absence of context-dependent effects.

\paragraph{\textbf{Simulation Details}}
For Table \ref{Tbl:SSAccuracy}(a), we performed two simulations for CDR3 sequences and eight simulations for full-length BCR sequences. For Table \ref{Tbl:SSAccuracy}(b), we conducted ten simulations for full-length BCR sequences. Evaluating the weight of a single tree on CDR3 sequences using the SMC algorithm with 1024 particles and 32 intermediate distributions requires approximately 12 hours on 45 CPU cores. Holding all other configurations fixed—including the tree topology and branch lengths—the CPU time increases approximately linearly with sequence length. For longer sequences, however, achieving stable estimates may require additional intermediate distributions and a larger number of particles, which can further increase the overall running time.

\begin{figure}[htbp]
    \centering
    \includegraphics[width=0.5\linewidth]{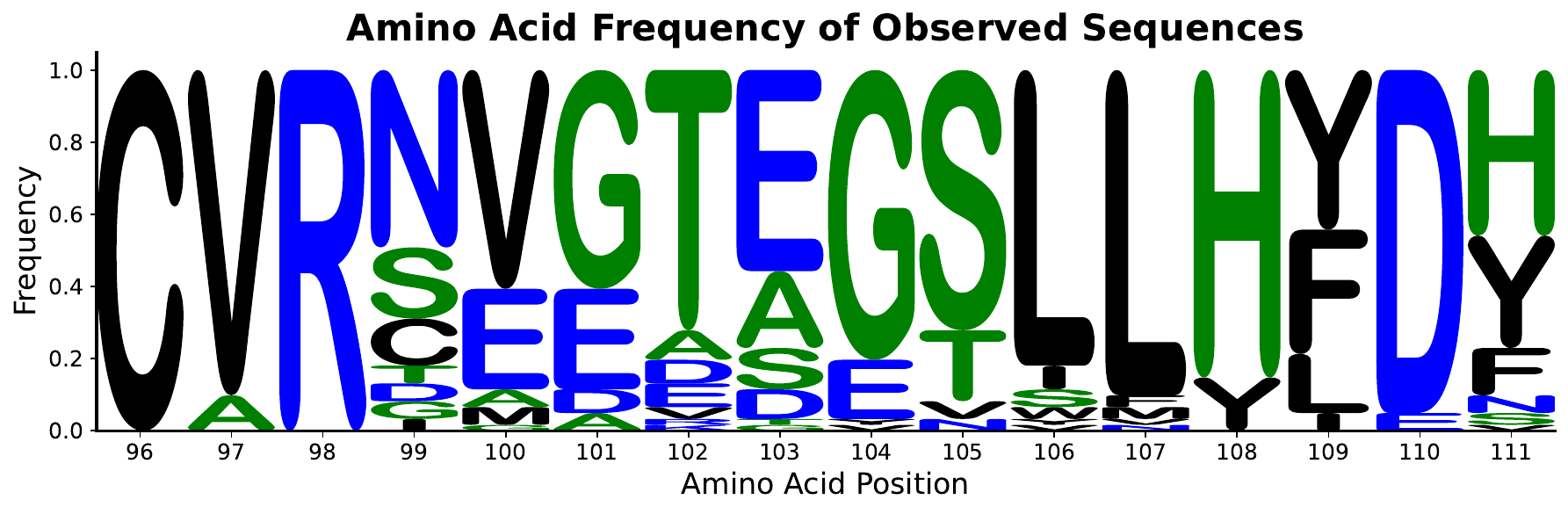}
    \caption{Frequency of observed sequences in CH235 clone.}
    \label{fig:CH235_observed}
\end{figure}

\begin{figure}
    \centering
    \begin{subfigure}[b]{0.85\linewidth}
        \centering
        \includegraphics[width=\linewidth]{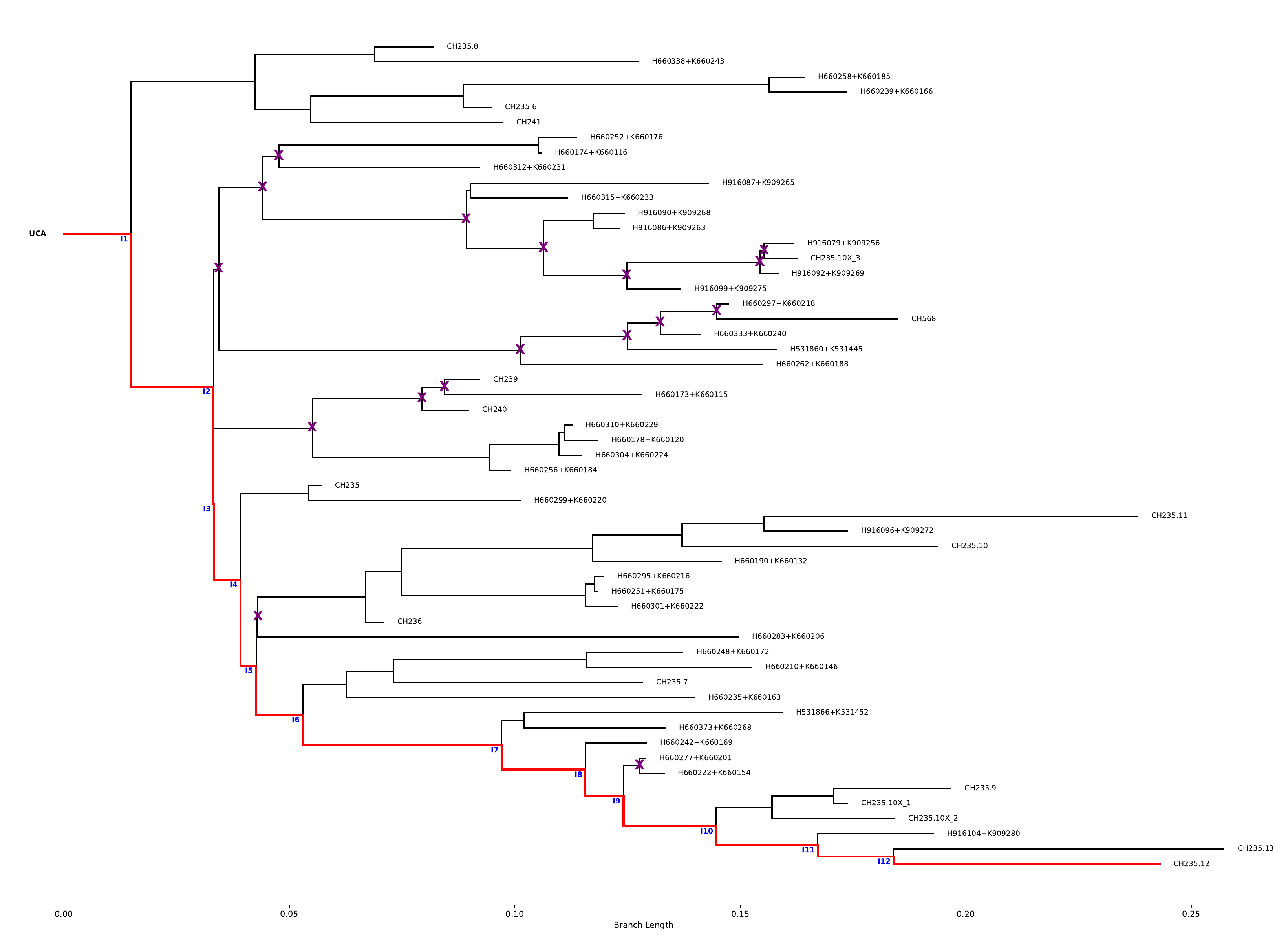}
        \caption{MAP CH235 clone under the ISM+NGS.}
        \label{fig:CH235_ISM_NGS}
    \end{subfigure}
    \\
    \begin{subfigure}[b]{0.85\linewidth}
        \centering
        \includegraphics[width=\linewidth]{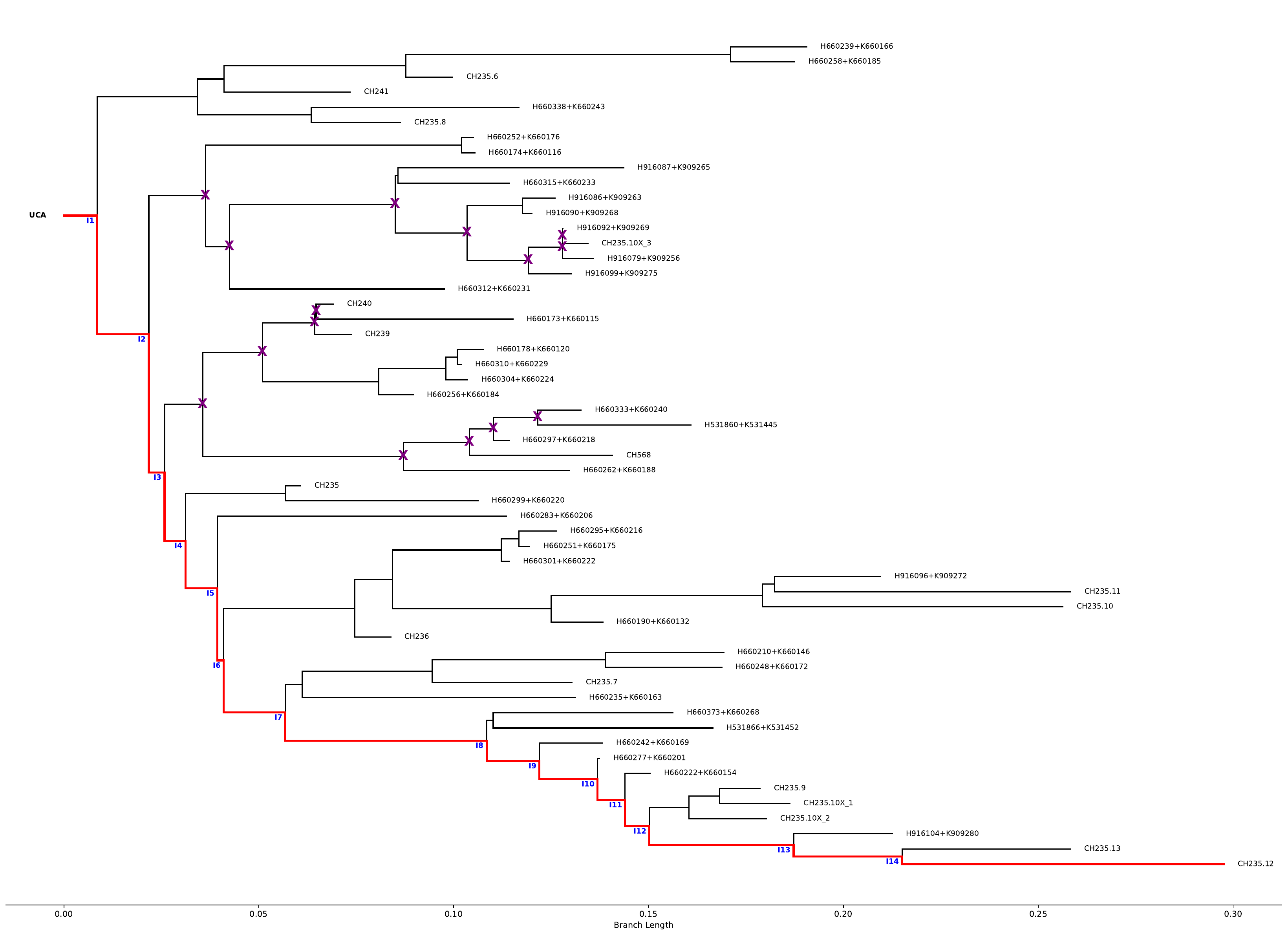}
        \caption{MAP CH235 clone under the DSM+NGS.}
        \label{fig:CH235_DSM_NGS}
    \end{subfigure}
    \caption{The CH235 lineage includes several antibodies, notably CH235, CH235.9, and CH235.12, each representing different stages of somatic hypermutation and maturation. Among these, CH235.12 is recognized as the most mature and broad member, capable of neutralizing approximately 90\% of circulating HIV-1 strains with high potency. This antibody has been extensively studied for its potential as a template for vaccine design. The red paths highlight the trajectory leading to CH235.12, and nodes marked with an `X' indicate subclones where the two trees differ in their inferred structures.}
    \label{fig:CH235_NGS}
\end{figure}

\paragraph{\textbf{CH235}} The amino acid frequencies in the CDR3 region of the 55 clonal members used in the experiments are shown in Figure~\ref{fig:CH235_observed}. The reconstructed CH235 clone under both the ISM and DSM frameworks are presented in Figure~\ref{fig:CH235_NGS}.

\end{document}